\documentclass[showpacs,nofootinbib,showkeys,eqsecnum,prd,aps,preprint]{revtex4}

\usepackage[english]{babel}
\usepackage[cp1251]{inputenc}
\usepackage[T1]{fontenc}

\usepackage{amssymb}
\usepackage{amsmath}
\usepackage{amsfonts}
\usepackage{latexsym}
\usepackage{mathrsfs}
\usepackage{bm}
\usepackage{tensor}
\usepackage{hyperref}

\begin{document}
	
\title{Relativistic quantum mechanics of massive neutrinos in a rotating frame}

\author{Alexander Breev}
\email{breev@izmiran.ru}

\author{Maxim Dvornikov}
\email{maxim.dvornikov@gmail.com}

\affiliation{Pushkov Institute of Terrestrial Magnetism, Ionosphere and Radiowave Propagation (IZMIRAN), 108840 Moscow, Troitsk, Russia}

\begin{abstract}
We study the evolution of neutrinos electroweakly interacting with a rotating matter. The description of neutrinos is based on the Dirac equation in the corotating noninertial frame where matter is at rest. We find solution of this Dirac equation, where the matter angular velocity is accounted for exactly, for massless neutrinos. In case of massive particles, this solution is obtained for a slowly rotating matter. Our findings are compared with previous research. We consider two applications of our results. First, we compute the electroweak contribution to the vector current of neutrinos along the rotation axis, which is analogous to the chiral vortical effect. This current is shown to be nonzero for both massless and massive particles. Then, we take into account the nonzero mixing between different mass eigenstates. It allows us to study neutrino flavor oscillations in rotating matter and account for noninertial effects. We derive the transition probability which reveals the resonance. These findings generalize the description of neutrino oscillations in a nonmoving matter. Some astrophysical applications are briefly discussed.
\end{abstract}



\maketitle

\section{Introduction\label{sec:INTR}}

Neutrinos were experimentally established, e.g., in Refs.~\cite{Fuk98,Ahm02} to have nonzero masses and mixing between different neutrino flavors. These neutrino properties lead to transitions between the neutrino flavor states called neutrino flavor oscillations. Despite neutrino flavor oscillations can happen in vacuum, the neutrino interaction with external fields significantly modifies the dynamics of oscillations. The electroweak interaction of neutrinos with background matter is one of the most important from the phenomenological point of view. Under certain circumstances, it can result in the significant amplification of the transition probability of neutrino oscillations called the Mikheyev--Smirnov--Wolfenstein (MSW) effect~\cite{Wol78,MikSmi85}. The MSW effect is likely to be the solution of the solar neutrino problem~\cite{CheXu25}.

The impact of external fields on oscillations is manifested more clearly for astrophysical neutrinos. Supernovae (SN) are known to be among the brightest neutrino sources in the universe. A core collapsing SN is believed to emit almost all its gravitational energy in the form of neutrinos (see, e.g., Ref.~\cite{GiuKim07}). Besides neutrino oscillations, various interesting phenomena, can take place in such dense neutrino fluxes. If one has even a small asymmetry in the neutrino emission in a certain direction, then, a protoneutron star receives a kick in the opposite direction because of the total momentum conservation. Such an effect can explain great linear velocities of some pulsars~\cite{Lor97}. Moreover, the correlation between the pulsar velocity and its spin was reported in Ref.~\cite{Joh05}.

The mechanism of the asymmetric neutrino emission by pulsars possessing a strong magnetic field was considered, e.g., in Refs.~\cite{KusSer96,LaiQia98}. Recently, the asymmetry in the fluxes of astrophysical neutrinos were studied in the context of various chiral phenomena. Here, we mention the chiral magnetic effect~\cite{Vil80,NieNin83}, which is the generation of a vector current along an external magnetic field. The chiral vortical effect (CVE) is more important for neutrinos since they are electrically neutral particles. The CVE is the generation of a vector or an axial current along the matter vorticity~\cite{SonSur09}. The application of chiral phenomena for SN neutrinos was considered in Refs.~\cite{Kam16,YamYan23}.

Formally, chiral phenomena disappear if particles are massive~\cite{Dvo16,Dvo18a,Dvo18b}. It is interesting to track the dependence of the currents generation on the particle masses. We studied the chiral effects of astrophysical neutrinos in Ref.~\cite{Dvo15}. The main tool in our previous works~\cite{Dvo15,Dvo14} was the exact solution of the Dirac equation in a rotating background matter. Solving this wave equation for massive neutrinos is quite challenging. Nevertheless, if one is interested in the study of the neutrino mass impact on, e.g., the CVE, such a solution is extremely useful. We also mention that this kind of Dirac equation was analyzed in Refs.~\cite{GriSavStu07,BalPopStu11}. The description of spinning particles in noninertial frames, based on the Dirac equation method, was reviewed in Ref.~\cite{Ver23}.

In the present work, we revisit our previous achievements in solving the Dirac equation for massive neutrinos in rotating matter. We notice that the neutrino mass was accounted for in Refs.~\cite{Dvo14,GriSavStu07} perturbatively. Now, we look for the solution of the Dirac equation in a systematic way by considering both massless and massive particles.

The reason for the revisiting of Dirac equation in a rotating frame is the following. While solving the Dirac equation in an field, one usually `squares' it to get rid of the Dirac matrices (see, e.g., Ref.~\cite{BagGit14book}). The electroweak interaction with matter involves the $\gamma^5$ matrix, see Sec.~\ref{sec:NUMIXMATT}. It makes the `squaring' to be peculiar. Moreover, the Dirac equation in question turns out to possess the nontrivial integral of motion which is found in our work.

Besides chiral phenomena, the exact solution of the Dirac equation is of importance for the description of neutrino flavor oscillations. Neutrino oscillations are known to disappear for massless particles. In Ref.~\cite{Dvo11}, we developed an approach, based on the relativistic quantum mechanics, for neutrino oscillations in various external fields, which involved exact solutions of the Dirac equations. Therefore, we apply the results of the present work to the description of neutrino flavor oscillations in rotating matter.

Our work is organized in the following way. First, in Sec.~\ref{sec:NUMIXMATT}, we recall some basic issues of the neutrino masses and mixing, as well as how flavor neutrinos can interact the background matter. We write down the Dirac equation for a neutrino interacting with background matter in a corotating frame in Sec.~\ref{sec:DIRACEQSOL}. In Sec.~\ref{sec:DIRACEQSOL}, a new method is proposed for squaring the Dirac equation for massless neutrinos, enabling the construction of a complete set of solutions without assuming a small rotation speed. The results of Sec.~\ref{sec:DIRACEQSOL} are applied in Sec.~\ref{sec:INDCURR} to calculate the induced vector current along the matter vorticity. 

In Sec.~\ref{sec:MASSSLOWROT}, using the transformation in Eq.~(\ref{8.1}), we derive the effective Dirac Eq.~(\ref{8.2b}) for neutrinos in a slowly rotating matter. We identify a nontrivial second-order symmetry operator, enabling the construction of a complete set of exact solutions for massive neutrinos therein. We are mainly interested in the description of neutrino flavor oscillations in rotating matter in Sec.~\ref{sec:NUOSC}. However, we also consider the contribution to the current for massive particles in Sec.~\ref{sec:CURRMASS}. Finally, we conclude in Sec.~\ref{sec:CONCL}.

Some details for writing down the Dirac equation in a noninertial frame are provided in Appendix~\ref{sec:ROTFRAME}. The useful properties the Laguerre functions are summarized in Appendix~\ref{sec:LAGPROP}. The mathematical details for the energy spectrum of massive neutrino  obtaining can be found in Appendix~\ref{sec:ENLEV}.

\section{Neutrino interaction with background matter\label{sec:NUMIXMATT}}

In this section, we remind how neutrinos can interact with background matter in Minkowski space. Here, the Dirac matrices have values corresponding to the flat space-time.

We consider the system of active flavor neutrinos, $\nu_\lambda = \nu_e, \nu_\mu,\dots$. These particles interact with other fermions in the standard model. However, flavor neutrinos do not have definite masses. The most general Lagrangian for the neutrino flavor eigenstates involves both Dirac and Majorana mass terms. Here, we assume that only the Dirac mass term is present
\begin{equation}\label{eq:masstermgen}
  - \mathcal{L}_m = \sum_{\lambda\lambda'} m_{\lambda\lambda'} \bar{\nu}_{\mathrm{R}\lambda} \nu_{\mathrm{L}\lambda}
  + \text{h.c.},
\end{equation}
where $(m_{\lambda\lambda'})$ is the nondiagonal mass matrix and $\nu_{\mathrm{R\lambda,L\lambda}}$ are the chiral projections.

The mass term in Eq.~\eqref{eq:masstermgen} can be diagonalized with help of the matrix transformation 
\begin{equation}\label{eq:massstates}
  \nu_{\lambda}=\sum_{a}U_{\lambda a}\psi_{a},
\end{equation}
where the fields $\psi_a$, $a = 1,2,\dots$, are called mass eigenstates since they possess definite masses $m_a$. The Lagrangian for $\psi_a$, written in flat space-time, takes the form,
\begin{equation}\label{eq:masstermgen}
  \mathcal{L} = \sum_{a} \bar{\psi}_a (\mathrm{i}\gamma^\mu \partial_\mu - m_a) \psi_a,
\end{equation}
where $\gamma^\mu = (\gamma^0,\bm{\gamma})$ are the Dirac matrices. The mass eigenstates $\psi_a$ are Dirac particles.

To describe the interaction of flavor neutrinos with background matter we use the Fermi model in the forward scattering approximation. In this case, the matter interaction Lagrangian has the form,
\begin{equation}\label{eq:mattintgen}
  \mathcal{L}_\mathrm{int} =
  \frac{1}{2}
  \sum_{\lambda}
  f_{\lambda}^{\mu}
  \bar{\nu}_{\lambda}
  \gamma_\mu
  (1-\gamma^5)
  \nu_{\lambda},
\end{equation}
where $\gamma^5 = \mathrm{i} \gamma^0\gamma^1\gamma^2\gamma^3$. The effective potential $f_{\lambda}^{\mu}$ in background matter, consisting of electrons, protons, and neutrons, reads
\begin{equation}\label{eq:effpotgen}
  f_{\lambda}^{\mu}
  =
  \sum_{f = e,p,n}
  (q_f^{(1)} j_f^\mu + q_f^{(2)} \lambda_f^\mu),
\end{equation}
where $j_f^\mu$ is the hydrodynamic current and $ \lambda_f^\mu$ is the four vector of polarization. The constants $q_f^{(1,2)}$ for different neutrino flavors can be found in the explicit form in Ref.~\cite{DvoStu02}.

One can see in Eq.~\eqref{eq:mattintgen} that the matter interaction potential in diagonal in the neutrino flavors. It is the feature of the standard model. However, if we use Eq.~\eqref{eq:massstates} to transform Eq.~\eqref{eq:mattintgen} to the mass basis, the total Lagrangian reads,
\begin{equation}\label{eq:Lagrmattmass}
	\mathcal{L}=\sum_{a,b=1,2}
	\bar{\psi}_{a}
	\left[
	\delta_{ab}(\mathrm{i}\gamma^{\mu}\partial_\mu - m_{a})
	-\frac{1}{2}g_{ab}^{\mu}\left(1-\gamma^{5}\right)
	\right]\psi_{b},
\end{equation}
where 
\begin{equation}\label{eq:mattpotmass}
	g_{ab}^{\mu}=\sum_{\lambda=e,\mu}U_{\lambda a}^{\dag}f_{\lambda}^{\mu}U_{\lambda b},
\end{equation}
is the nondiagonal matter potential in the mass basis. In Eqs.~\eqref{eq:Lagrmattmass} and~\eqref{eq:mattpotmass}, we consider the situation of two mass eigenstates.

\section{Dirac equation for neutrino in a noninertial frame\label{sec:DIRACEQSOL}}

In this section, we find the solution of the Dirac equation for a single mass eigenstate interacting with matter moving with an acceleration. For the sake of brevity, we replace $g_{ab}^{\mu} \to g^{\mu}$ in Eq.~\eqref{eq:mattpotmass}. We consider the evolution of a neutrino in a comoving frame.

Let us study the situation when the background matter rotates with the constant angular velocity $\omega$. For simplicity, we take that one has no differential rotation between background fermions of the different type. Moreover, we assume that the background matter is unpolarized.

We describe the dynamics of the system in the corotating frame. In this situation, one has the only nonzero component of the vector $g^\mu$, $g^0 \neq 0$. In general case, $g^\mu \propto j^\mu = n_0 u^\mu$, where $n_0$ is the invariant number density and $u^\mu$ is the four velocity. Thus, $g^0 \propto n_0 \tfrac{\mathrm{d}t}{\mathrm{d}s} = n$, where $n$ is the number density measured by an observer at rest. We assume that both $n$ and $g^0$ are constant.

The local coordinates of a nonrotating observer in a four-dimensional Minkowski space $\mathscr{M}$ are $(x^{0},x^{1},x^{1},x^{3})$. The metric tensor
on $\mathscr{M}$ is $(\eta_{ab})=\mathrm{diag}(1,-1,-1,-1)$, with $a,b=0,\dots,3$. The interval of $\mathscr{M}$ in cylindrical coordinates $\{t',r',\phi',z'\}$
has the form ($x^{0}=t',x^{1}=r'\cos\phi',x^{2}=r'\sin\phi',x^{3}=z'$),
\begin{equation}
ds^{2}=dt'{}^{2}-dr'{}^{2}-r'{}^{2}d\varphi'{}^{2}-dz'{}^{2}\,.\label{1.1}
\end{equation}

We denote the nee rotating cylindrical coordinates as $x=\left(x^{\mu}\right)=\left(t,r,\phi,z\right)$
and assume that the axis of rotation is along $z$ and $z'$:
\begin{equation}
t'=t,\quad r'=r,\quad z'=z,\quad\phi'=\phi-\omega t\,,\label{1.2}
\end{equation}
where $\omega$ is the angular velocity of the reference frame. Then,
the interval $ds^{2}$ in this coordinate system takes the form
\begin{equation}
 ds^{2}=g_{\mu\nu}(x)dx^{\mu}dx^{\nu}=\left(1-\omega^{2}r^{2}\right)dt^{2}-dr^{2}-2\omega r^{2}dtd\phi-r^{2}d\phi^{2}-dz^{2}\,,
 \label{1.3}
\end{equation}

The Dirac equation for a neutrino mass eigenstate in a curved space-time can be obtained on the basis of Eq.~\eqref{eq:masstermgen} if we replace $\gamma^\mu \to \gamma^\mu(x)$ and $\partial_\mu \to \nabla_{\mu}+\Gamma_{\mu}$, where $\gamma^\mu(x)$ are the coordinate dependent Dirac matrices, $\nabla_{\mu}$ is the covariant derivative, and $\Gamma_{\mu}$ is the spin connection. (For detailed information on the Dirac equation in curved spacetime, we refer the reader to Refs.~\cite{Collas2019,Birrell1984, BagGit90book, BagGit14book}). Using the results of Appendix~\ref{sec:ROTFRAME}, we get the Dirac operator $\mathscr{D}=i\gamma^{\mu}(x)\left(\nabla_{\mu}+\Gamma_{\mu}\right)$, which acts on the wavefunction $\psi(t,r,\phi,z)$,
in the metric in Eq.~(\ref{1.3}) as
\begin{align}
\mathscr{D}(x,\partial_{x}) & =i\gamma^{0}\partial_{t}+i\gamma^{1}\left(\partial_{r}+\frac{1}{2r}\right)-i\left(\omega\gamma^{0}-\frac{1}{r}\gamma^{2}\right)\partial_{\phi}+\text{i\ensuremath{\gamma}}^{3}\partial_{z}\,.\label{2.6}
\end{align}

We will consider the motion of a single massive neutrino in matter rotating with
a constant angular velocity $\omega$ in the rotating frame. Thus, we replace $m_a \to m$. In this case, the Dirac equation takes the form,
\begin{equation}
\left[\mathscr{D}(x,\partial_{x})-m\right]\psi(x)=\frac{1}{2}\gamma_{0}(x)g^{0}\left[1-\gamma^{5}(x)\right]\psi(x)\,,\label{3.1}
\end{equation}
where $\gamma_{0}(x)=\gamma^{0}-\omega r\gamma^{2}$, $\gamma^{5}(x)=\gamma^{5}$.

Equation~(\ref{1.3}) admits three commuting integrals of motion
\begin{equation}
\hat{p}_{0}=i\partial_{t},\quad\hat{p}_{z}=-i\partial_{z},\quad\hat{J}_{z}=-i\partial_{\phi}\,.\label{3.2}
\end{equation}
Let us consider the eigenvalue problem,
\begin{equation}
\hat{p}_{0}\psi=E\psi,\quad\hat{p}_{z}\psi=p_{z}\psi,\quad\hat{J}_{z}\psi=J_{z}\psi\,.\label{3.3}
\end{equation}
Accounting for Eq.~(\ref{3.3}), one gets the spinor wavefunction in the form,
\begin{equation}
\psi=\psi_{w}(x)=\psi_{w}(r)\exp\left(-iEt+iJ_{z}\phi+ip_{z}z\right),\quad w=\{E,J_{z},p_{z}\}\,.\label{3.4}
\end{equation}

Based on the periodic boundary conditions,
\begin{equation}
\left.\psi_{w}(x)\right|_{\phi=\tilde{\phi}+2\pi}=-\left.\psi_{w}(x)\right|_{\phi=\tilde{\phi}},\quad\tilde{\phi}\in[0;2\pi)\,,\label{3.4b}
\end{equation}
one gets that
\begin{equation}
J_{z}=\frac{1}{2}-\ell,\quad\ell\in\mathbb{Z},\quad w=\{E,\ell,p_{z}\},\quad\mathbb{Z}=\left\{ 0,\pm1,\pm2,\dots\right\} \,.\label{3.4c}
\end{equation}
 
Substituting Eq.~(\ref{3.4}) into Eq.~(\ref{3.1}), we obtain the reduced
Dirac equation for the spinor $\psi_{n}(r)$:
\begin{align}
 & \left[i\gamma^{1}\left(  \frac{d}{dr}+\frac{1}{2r}\right)+\gamma^{0}\left(E+\omega J_{z}-\frac{g^{0}}{2}\right)-\gamma^{2}\left(\frac{J_{z}}{r}-\frac{g^{0}}{2}\omega r\right)\right.\nonumber \\
 & \,\,\,-\left.p_{z}\gamma^{3}-m+\frac{g^{0}}{2}\left(\gamma^{0}-\omega r\gamma^{2}\right)\gamma^{5}\right]\psi_{w}(r)=0\,,\label{3.5}
\end{align}
which depends on radial coordinate only.

First, we look for the complete set of solutions of Eq.~(\ref{3.5}) for ultrarelativistic
neutrinos if we neglect the particle mass. For this purpose, we represent these solutions in the form,
\begin{align}
\psi_{w}(r)= & \left[i\gamma^{1}\left(\frac{d}{dr}+\frac{1}{2r}\right)+\gamma^{0}\left(E+\omega J_{z}-\frac{g^{0}}{2}\right)-\gamma^{2}\left(\frac{J_{z}}{r}-\frac{g^{0}}{2}\omega r\right)\right.\nonumber \\
 & \left.\,\,\,-p_{z}\gamma^{3}-\frac{g^{0}}{2}\left(\gamma^{0}-\omega r\gamma^{2}\right)\gamma^{5}\right]\Phi_{w}(r)\,,\label{4.1}
\end{align}
where $\Phi_{w}(r)$ is the new auxiliary spinor. Unlike the traditional method of squaring the Dirac equation (see, e.g., Refs.~\cite{BagOb92,BagGit90book,BagGit14book}), in Eq.~(\ref{4.1}), we change the sign of the terms containing the matrix $\gamma^{5}$. This approach is justified by the fact that, as a result of squaring, namely, the substituting Eq.~(\ref{4.1}) into Eq.~(\ref{3.5}), we obtain an equation for the spinor $\Phi_{w}(r)$ that does not contain matrix coefficients at the derivatives,
\begin{align}
 & \mathscr{D}^{2}(r)\Phi_{w}(r)=0,\nonumber \\
 & \mathscr{D}^{2}(r)=\frac{d^2}{dr^2}+\frac{1}{r}\frac{d}{dr}+i\left(\frac{J_{z}}{r^{2}}+\frac{\omega g^{0}}{2}\right)\gamma^{12}+\frac{\omega g^{0}}{2}\gamma^{03}-g^{0}\left[E-\frac{g^{0}}{2}\left(1-\omega^{2}r^{2}\right)\right]\gamma^{5}\nonumber \\
 & -\left(J_{z}^{2}+\frac{1}{4}\right)\frac{1}{r^{2}}+\left(E+\omega J_{z}\right)^{2}+\left(g^{0}\right)^{2}\left(1-\frac{1}{2}\omega^{2}r^{2}\right)-\frac{1}{2}\left[\left(E+g^{0}\right)^{2}-E^{2}\right]-p_{z}^{2}.\label{4.2}
\end{align}
One can see that Hermitian matrices,
\[
Q_{1}=Q_{1}^{\dag}=i\gamma^{12},\quad Q_{2}=Q_{2}^{\dag}=\gamma^{5},\quad\left[Q_{1},Q_{2}\right]=0,\quad Q_{k}^{2}=1\,.
\]
commute with the operator $\mathscr{D}^{2}$. Thus, they are the symmetry operators of Eq.~(\ref{4.2}).

We look for the solutions of Eq.~(\ref{4.2}) in the class of functions that are eigenfunctions of the matrices $Q_{k}$ and correspond to the eigenvalues $\chi^{2}=1$ and $\sigma^{2}=1$. Then
\begin{align}
\Phi_{w}(r) & =\Phi_{w,\sigma,\chi}(r)=\varphi_{w,\sigma,\chi}(r)U_{\sigma,\chi},\label{4.2b}\\
Q_{1}U_{\sigma,\chi} & =\chi U_{\sigma,\chi},\quad Q_{2}U_{\sigma,\chi}=\sigma U_{\sigma,\chi}\,,\nonumber 
\end{align}
where $\varphi_{w,\sigma,\chi}(r)$ are  scalar functions; $U_{\sigma,\chi}$ are constant spinors, with respect to the representation in Eq.~(\ref{2.1b}) having the form: 
\begin{align}
 & U_{1,1}=\begin{pmatrix}1\\
0\\
0\\
0
\end{pmatrix},\quad U_{1,-1}=\begin{pmatrix}0\\
1\\
0\\
0
\end{pmatrix},\quad U_{-1,1}=\begin{pmatrix}0\\
0\\
-1\\
0
\end{pmatrix},\quad U_{-1,-1}=\begin{pmatrix}0\\
0\\
0\\
-1
\end{pmatrix}\,,\label{4.2c}\\
 & U_{\sigma,\chi}^{\dag}U_{\sigma,'\chi'}=\delta_{\chi,\chi'}\delta_{\sigma,\sigma'}\,.
\end{align}

Substituting Eq.~(\ref{4.2b}) into Eq.~(\ref{4.2}), we have an ordinary differential equation for the function $\varphi_{w,\sigma,\chi}(r)$:
\begin{align}
 & \mathscr{L}_{\ell,\sigma,\chi}(r)\varphi_{w,\sigma,\chi}(r)\nonumber \\
 & =\varphi_{w,\sigma,\chi}''(r)+\frac{1}{r}\varphi_{w,\sigma,\chi}'(r)+\left\{ \left(E+\omega J_{z}\right)^{2}-p_{z}^{2}-\frac{1}{r^{2}}\left(J_{z}-\frac{\chi}{2}\right)^{2}\right.\nonumber \\
 & \left.+\frac{\sigma+1}{2}g^{0}\left[\chi\omega-2E+g^{0}\left(1-\omega^{2}r^{2}\right)\right]\right\} \varphi_{w,\sigma,\chi}(r)=0\,,\nonumber\\
 &\varphi''(r):=\frac{d^2\varphi(r)}{dr^2},\quad \varphi'(r):=\frac{d\varphi(r)}{dr}\,.\label{4.6c}
\end{align}

First, we consider the case $\sigma=1$. It corresponds to active neutrinos. We denote $E$ for $\sigma=1$ as $E_{A}$, and the solution of Eq.~(\ref{4.6c}) as $\varphi_{w,\chi}(r):=\varphi_{w,\ell,\sigma=1,\chi}(r)$. With respect to the new independent variable $\rho=\omega\left|g^{0}\right|r^{2}$, Eq.~(\ref{4.6c}) is reduced to a differential equation on the Laguerre function $I_{u,v}(\rho)$:
\begin{align}
 & 4\rho^{2}I''_{u,v}(\rho)+4\rho I'_{u,v}(\rho)-\left[\rho^{2}-2\rho\left(u+v+1\right)+\left(v-u\right)^{2}\right]I_{u,v}(\rho)=0\,,\nonumber \\
 & \alpha=u-v=-\left(J_{z}-\frac{\chi}{2}\right)=\ell-\frac{1-\chi}{2}\in\mathbb{Z},\quad u=\lambda-\frac{1}{2}\left(J_{z}-\frac{\chi}{2}\right)-\frac{1}{2}\,,\label{4.16}
\end{align}
where
\[
\lambda=\frac{1}{4\omega\left|g^{0}\right|}\left[\left(E_{A}+\omega J_{z}\right)^{2}-\left(E_{A}-\frac{\chi\omega}{2}\right)^{2}-p_{z}^{2}+\left(E-\frac{\chi\omega}{2}-g^{0}\right)^{2}\right]\,.
\]
The general solution of Eq.~(\ref{4.6c}) is represented as
\begin{equation}
\varphi_{w,\chi}(r)=C_{1}I_{u,v}\left(\omega\left|g^{0}\right|r^{2}\right)+C_{2}I_{v,u}\left(\omega\left|g^{0}\right|r^{2}\right),\quad C_{k}=\mathrm{const}\,,\label{5.5a}
\end{equation}
where $I_{u,v}$ is the Laguerre function. We provide some of the properties of Laguerre functions in Appendix~\ref{sec:LAGPROP}. In Eq.~\eqref{4.16}, the integer parameter $\left|\alpha\right|\geq1$, or $\alpha=0$: 
\begin{align}
 & \left|\alpha\right|=\left|\ell-\frac{1-\chi}{2}\right|\geq1,\quad\ell\neq\frac{1-\chi}{2},\nonumber \\
 & \alpha=0,\quad\ell=\frac{1-\chi}{2}\,.\label{4.16b}
\end{align}

We define the complete set of solutions of Eq.~(\ref{4.6c}) as follows:
\begin{equation}
\varphi_{w,\chi}(r)=\sqrt{2\left|g^{0}\right|\omega}\begin{cases}
I_{u,v}(\rho), & g^{0}>0,\\
I_{v,u}(\rho) & g^{0}<0.
\end{cases}\,\label{4.16bb}
\end{equation}
Since $\left|u-v\right|=\left|\alpha\right|\in\mathbb{Z}_{+}$, the quadratic integrability condition
\begin{equation}
\int_{0}^{+\infty}\left|\varphi_{w,\chi}(\rho)\right|^{2}d\rho<\infty\label{4.17}
\end{equation}
can be ensured if
\begin{align}
 & v=s\in\mathbb{Z}_{+},\quad u=v+\alpha=s+\alpha=s+\ell-\frac{1-\chi}{2}=n\in\mathbb{Z}_{+},\quad\textrm{if}\quad g^{0}>0\,,\label{4.17a}\\
 & u=s\in\mathbb{Z}_{+},\quad v=u-\alpha=s-\alpha=s-\ell+\frac{1-\chi}{2}=n\in\mathbb{Z}_{+},\quad\textrm{if}\quad g^{0}<0\,.
\end{align}
Then, instead of the quantum number $\ell$, we consider the new quantum number
\begin{equation}
s=n-\mathrm{sign}\left(g^{0}\right)\ell+(1-\chi)/2,\label{4.17aa}
\end{equation}
and number the bounded solutions (decreasing at infinity) of Eq.~(\ref{4.16bb}) using the set of quantum numbers $n$ and $s$:
\begin{align}
 & \varphi_{w,\chi}(r)=\varphi_{n,s}(r)=\sqrt{2\left|g^{0}\right|\omega}I_{n,s}\left(\omega\left|g^{0}\right|r^{2}\right),\quad\lambda=\lambda_{n}=n+\frac{1-\mathrm{sign}\left(g^{0}\right)\alpha}{2}\,,\nonumber \\
 & I_{n,s}\left(\rho\right)=\sqrt{\frac{n!}{s!}}\rho^{\left(n-s\right)/2}e^{-\rho/2}L_{s}^{n-s}(\rho)\,,\label{4.17b}
\end{align}
Using the properties of the Laguerre functions in Appendix~\ref{sec:LAGPROP}, one gets that the set of solutions in Eq.~(\ref{4.17b}) satisfies the normalization and completeness conditions:
\begin{eqnarray}
	\int_{0}^{\infty}\left|\varphi_{n,s}(r)\right|^{2}rdr&=&1\,,
	\label{4.20a}\\
	\sum_{s=0}^{\infty}\varphi_{n,s}(r)\varphi_{n,s}(r') &=& \frac{1}{r}\delta(r-r')\,. 
	\label{4.20aa}
\end{eqnarray}

Substituting Eq.~(\ref{4.16}) into Eq.~(\ref{4.17b}), we obtain a condition on $\lambda$ that is equivalent to satisfying the condition of quadratic integrability:
\begin{align}
 & \left(E_{A}-g^{0}\right)^{2}+\left(E_{A}+\omega J_{z}\right)^{2}-E_{A}^{2}=p_{z}^{2}+4\omega g^{0}\text{\ensuremath{\left[n_{\chi}+\mathrm{sign}\left(g^{0}\right)\frac{J_{z}}{2}\right]}}\,,\label{4.22-1}\\
 & n_{\chi}:=n+\frac{1-\mathrm{sign}\left(g^{0}\right)\chi}{2}\,.
\end{align}
From Eq.~(\ref{4.22-1}) we obtain an expression for the neutrino spectrum:
\begin{equation}
E_{A}=E_{n_{\chi}}=g^{0}-\omega J_{z}+\zeta\mathscr{E}_{n_{\chi}},\quad\mathscr{E}_{n_{\chi}}=\sqrt{4\left|g^{0}\right|\omega n_{\chi}+p_{z}^{2}},\quad\zeta=\pm1\,,\label{4.22-2}
\end{equation}
where the parameter $\zeta=+1$ corresponds to particles, and $\zeta=-1$ to antiparticles.

Condition in Eq.~(\ref{4.17a}) can be satisfied if and only if the quantum number $\ell$ satisfies the inequality,
\begin{equation}
-\infty<\mathrm{sign}\left(g^{0}\right)\left(\ell-\frac{1-\chi}{2}\right)\leq n\,.\label{4.19f}
\end{equation}
Indeed, assume that Eq.~(\ref{4.19f}) holds true. Then, $s=n-\mathrm{sign}\left(g^{0}\right)[\ell-(1-\chi)/2]\geq0$ and Eq.~(\ref{4.17a}) is valid. Note that, if condition in Eq.~(\ref{4.19f}) is violated, then bounded solutions of Eq.~(\ref{4.16}) can be given in the form,
\begin{align}
 & \varphi_{w,\chi}(r)=\varphi_{N,\ell}(r)=\sqrt{2\left|g^{0}\right|\omega}I_{N+\left|\alpha\right|,N}\left(\omega\left|g^{0}\right|r^{2}\right)\,,\nonumber\\
 &\lambda=\lambda_{N}=N+\frac{1+\left|\alpha\right|}{2},\quad N\in\mathbb{Z}_{+}\,.\label{4.19g}
\end{align}
In this case, the spectrum has the form in Eq.~(\ref{4.22-2}), where
\begin{align}
 & n_{\chi}=N_{J_{z},\chi}:=N+\frac{1}{2}\left[\left|J_{z}-\frac{\chi}{2}\right|-\left(\mathrm{sign}\left(g^{0}\right)J_{z}-\frac{\chi}{2}\right)+\frac{1-\frac{1+\mathrm{sign}\left(g^{0}\right)}{2}\chi}{2}\right]\,.\label{4.19gg}
\end{align}

Now, we consider the case $\sigma=-1$, which corresponds to the absence of interaction, since the operator of Eq.~(\ref{4.6c}) does not depend on $g^{0}$. We denote $E$ for $\sigma=-1$ as $E_{S}$. We define the solution of Eq.~(\ref{4.6c}) as $\varphi_{\mu,\ell,\chi}(r):=\varphi_{w,\ell,\sigma=-1,\chi}(r)$.

Equation~(\ref{4.6c}) in this case is reduced to an equation for the Bessel function:
\begin{align}
 & r^{2}\varphi_{\mu,\ell,\chi}''(r)+r\varphi_{\mu,\ell,\chi}'(r)+\left(k^{2}r^{2}-\alpha^{2}\right)\varphi_{\mu,\ell,\chi}(r)=0\,,\nonumber \\
 & \alpha=\left|J_{z}-\frac{\chi}{2}\right|=\left|\ell-\frac{1-\chi}{2}\right|\in\mathbb{Z}_{+}\,,\quad\left(E_{S}+\omega J_{z}\right)^{2}=k^{2}+p_{z}^{2}\,,\quad\mu>0\,.\label{4.23}
\end{align}
Equation (\ref{4.23}) has a general solution
\[
\varphi_{k,\ell,\chi}(r)=C_{1}J_{\alpha}\left(k r\right)+C_{2}N_{\alpha}\left(k r\right)\,,
\]
where $J_{\alpha}(k r)$ is the Bessel function of integer index $\alpha$,
\begin{equation}
\int_{0}^{\infty}J_{\alpha}(k r)J_{\alpha}(k'r)rdr=\frac{1}{k}\delta\left(k-k'\right)\,.\label{4.24}
\end{equation}
Since the Neumann function $N_{\nu}(k r)$ is not bounded at zero, we set $C_{2}=0$.

Thus, in the case $\sigma=-1$, we have a continuous spectrum,
\begin{equation}
E_{S}=\zeta\sqrt{k^{2}+p_{z}^{2}}-\omega J_{z},\quad\zeta=\pm1\,.\label{4.25}
\end{equation}
which is numbered by the parameter $k$.

\section{Induced current along the rotation axis\label{sec:INDCURR}}

In this section, we compute the current of neutrinos along the rotation axis. The computation is based on the solution of the Dirac equation found in Sec.~\ref{sec:DIRACEQSOL}.

The phenomenon described is analogous to the CVE, i.e. the generation of the vector current $\mathbf{J}$ along the vorticity of matter $\bm{\omega}$~\cite{Kha16}, $\mathbf{J} \propto \bm{\omega}$. Note that we do not consider the generation of the axial current $\mathbf{J}_5$ in rotating medium, which is also called the CVE~\cite{Kha16}.

\subsection{Complete set of solutions for the massless case}

The calculation of the current is based on the complete set of solutions of the Dirac equation for a neutrino in the rotating matter. Using Eq.~(\ref{4.1}), we construct a set of solutions of the reduced Dirac Eq.~(\ref{3.5}) for $m=0$ for $\sigma=+1$:
\begin{align}
 & \,^{\zeta}\psi_{r}^{\pm}(r)=\,^{\zeta}C_{\pm}\begin{pmatrix}0\\
\,^{\zeta}\psi_{\pm}
\end{pmatrix},\quad\,^{\zeta}\psi_{+}=\,^{\zeta}\Pi\begin{pmatrix}\varphi_{+}\\
0
\end{pmatrix},\quad\psi_{-}=\,^{\zeta}\Pi\begin{pmatrix}0\\
\varphi_{-}
\end{pmatrix},\quad\varphi_{\pm}=\varphi_{n,s,\chi=\pm1}(r)\,,\nonumber \\
 & \,^{\zeta}\Pi=\begin{pmatrix}p_{z}-\zeta\mathscr{E}_{n_{+}} & \sqrt{g^{0}\omega}R_{-}\\
\sqrt{g^{0}\omega}R_{+} & -\left(p_{z}-\zeta\mathscr{E}_{n_{-}}\right)
\end{pmatrix},\quad R_{\pm}=-i\sqrt{\rho}\left[2\frac{d}{d\rho}\pm\mathrm{sign}\left(g^{0}\right)\left(\frac{n-s}{\rho}+1\right)\right]\,,\nonumber \\
 & \,^{\zeta}C_{\pm}=\left[4\left|g^{0}\right|\omega n_{\pm}+\left(p_{z}\mp\zeta\mathscr{E}_{n_{\pm}}\right)^{2}\right]^{-1/2},\quad\mathscr{E}_{n_{\pm}}=\sqrt{4\left|g^{0}\right|\omega n_{\pm}+p_{z}^{2}}\,,\nonumber \\
 & n_{+}=\begin{cases}
n, & g^{0}>0,\\
n+1 & g^{0}<0.
\end{cases},\quad n_{-}=\begin{cases}
n+1, & g^{0}>0,\\
n & g^{0}<0.
\end{cases}\,.\label{6.1}
\end{align}

Considering that
\begin{equation}
R_{\pm}I_{n,s}(\rho)=\begin{cases}
\mp2i\sqrt{n_{\pm}}I_{n\mp1,s}(\rho), & g^{0}>0,\\
\pm2i\sqrt{n_{\pm}}I_{n\pm1,s}(\rho) & g^{0}<0,
\end{cases}\label{6.2}
\end{equation}
we obtain the set of solutions for $g^{0}>0$,
\begin{align}
\,^{\zeta}\psi_{+} & =\,^{\zeta}\psi_{n,s,\chi=1}(\rho)=\sqrt{2\left|g^{0}\right|\omega}\,^{\zeta}C_{+}\begin{pmatrix}\left(p_{z}-\zeta\mathscr{E}_{n}\right)I_{n,s}(\rho)\\
-2i\sqrt{\left|g^{0}\right|\omega n}I_{n-1,s}(\rho)
\end{pmatrix}\,,\nonumber \\
\,^{\zeta}\psi_{-} & =\,^{\zeta}\psi_{n,s,\chi=-1}(\rho)=\sqrt{2\left|g^{0}\right|\omega}\,^{\zeta}C_{-}\begin{pmatrix}2i\sqrt{\left|g^{0}\right|\omega(n+1)}I_{n+1,s}(\rho)\\
-\left(p_{z}+\zeta\mathscr{E}_{n}\right)I_{n,s}(\rho)
\end{pmatrix},\quad g^{0}>0\,,\label{6.3}
\end{align}
and for the case $g^{0}<0$,
\begin{align}
\,^{\zeta}\psi_{+} & =\,^{\zeta}\psi_{n,s,\chi=1}(\rho)=\sqrt{2\left|g^{0}\right|\omega}\,^{\zeta}C_{+}\begin{pmatrix}\left(p_{z}-\zeta\mathscr{E}_{n}\right)I_{n,s}(\rho)\\
2i\sqrt{\left|g^{0}\right|\omega(n+1)}I_{n+1,s}(\rho)
\end{pmatrix}\,,\nonumber \\
\,^{\zeta}\psi_{-} & =\,^{\zeta}\psi_{n,s,\chi=-1}(\rho)=\sqrt{2\left|g^{0}\right|\omega}\,^{\zeta}C_{-}\begin{pmatrix}-2i\sqrt{\left|g^{0}\right|\omega n}I_{n-1,s}(\rho)\\
-\left(p_{z}+\zeta\mathscr{E}_{n}\right)I_{n,s}(\rho)
\end{pmatrix},\quad g^{0}<0\,.\label{6.4}
\end{align}

As a result, we have the orthonormal set of solutions to the Dirac equation,
\begin{align}
 \,^{\zeta}\psi_{w}^{\pm}(x)&=\frac{1}{\sqrt{2\pi}}\begin{pmatrix}0\\
\,^{\zeta}\psi_{n,s,\chi=\pm1}\left(\omega\left|g^{0}\right|r^{2}\right)
\end{pmatrix}\times\nonumber\\
&\times\exp\left\{ -i\,^{\zeta}E_{n_{\pm}}t+i\left[\frac{\chi}{2}+\mathrm{sign}\left(g^{0}\right)\left(s-n\right)\right]\phi+ip_{z}z\right\} \,.\label{6.5}
\end{align}
The set of solutions in Eq.~(\ref{6.5}) satisfies the orthogonality condition
\begin{equation}
	\int\left[\,^{\zeta}\psi_{w}^{\pm}\left(x\right)\right]^{\dag}\,^{\zeta}\psi_{w'}^{\pm}\left(x\right)rdr=\delta_{w,w'}=\delta_{n,n'}\delta_{s,s'}\delta\left(p_{z}-p_{z}'\right)\,,
	\label{ort1}
\end{equation}
where $w=\{n,s,p_{z}\}$, $w'=\{n',s',p'_{z}\}$.

The matrix $Q_{1}=i\gamma^{12}$, although commuting with the operator $\mathscr{D}^{2}(r)$, does not correspond to the integral of motion of the Dirac equation. Therefore, the set of solutions $\psi_{w}^{+}(x)$ and $\psi_{w}^{-}(x)$ is linearly dependent. For $g^{0}>0$ we have:
\begin{align}
 & \sqrt{4\left|g^{0}\right|\omega n\left[4\left|g^{0}\right|\omega n+\left(p_{z}-\zeta\mathscr{E}_{n}\right)^{2}\right]}\,^{\zeta}\psi_{n,s,p_{z}}^{+}(x)\nonumber \\
 & +i\left(p_{z}-\zeta\mathscr{E}_{n}\right)\sqrt{4\left|g^{0}\right|\omega n+\left(p_{z}+\zeta\mathscr{E}_{n}\right)^{2}}\,^{\zeta}\psi_{n-1,s,p_{z}}^{-}(x)=0\,.\label{6.6}
\end{align}
For $g^{0}<0$ the identity is hold:
\begin{align}
 & \sqrt{4\left|g^{0}\right|\omega\left(n+1\right)\left[4\left|g^{0}\right|\omega\left(n+1\right)+\left(p_{z}-\zeta\mathscr{E}_{n+1}\right)^{2}\right]}\,^{\zeta}\psi_{n,s,p_{z}}^{+}(x)\nonumber \\
 & -i\left(p_{z}-\zeta\mathscr{E}_{n+1}\right)\sqrt{4\left|g^{0}\right|\omega\left(n+1\right)+\left(p_{z}+\zeta\mathscr{E}_{n+1}\right)^{2}}\,^{\zeta}\psi_{n+1,s,p_{z}}^{-}(x)=0\,.\label{6.7}
\end{align}
For this reason, the parameter $\chi=\pm1$ is not included in the set of quantum numbers $w$. Without loss of generality, we can fix this parameter. For convenience, we set $\chi=\mathrm{sign}(g^{0})$ and consider the set of solutions,
\begin{align}
\,^{\zeta}\psi_{w}(x) & =\,^{\zeta}\psi_{w}^{+}(x)=\frac{1}{\sqrt{2\pi}}\begin{pmatrix}0\\
\,^{\zeta}\psi_{n,s}\left(\omega g^{0}r^{2}\right)
\end{pmatrix}\exp\left\{ -i\,^{\zeta}E_{n}t+iJ_{z}\phi+ip_{z}z\right\} \,,\nonumber \\
\,^{\zeta}\psi_{n,s}(\rho) & =\sqrt{2\left|g^{0}\right|\omega}\,^{\zeta}C_{n,s}\begin{pmatrix}\left(p_{z}-\zeta\mathscr{E}_{n}\right)I_{n,s}(\rho)\\
-2i\sqrt{g^{0}\omega n}I_{n-1,s}(\rho)
\end{pmatrix},\quad g^{0}>0\,,\nonumber \\
\,^{\zeta}\psi_{n,s}(\rho) & =\sqrt{2\left|g^{0}\right|\omega}\,^{\zeta}C_{n,s}\begin{pmatrix}-2i\sqrt{g^{0}\omega n}I_{n-1,s}(\rho)\\
-\left(p_{z}+\zeta\mathscr{E}_{n}\right)I_{n,s}(\rho)
\end{pmatrix},\quad g^{0}<0\,,\nonumber \\
\,^{\zeta}C_{n,s} & =\left\{ 4\left|g^{0}\right|\omega n+\left[p_{z}-\zeta\mathscr{\mathrm{sign}\left(\textrm{\ensuremath{g^{0}}}\right)E}_{n}\right]^{2}\right\} ^{-1/2},\quad J_{z}=\mathrm{sign}\left(g^{0}\right)\left[s-n+\frac{1}{2}\right]\,.\label{6.8}
\end{align}
 
One can see in Eq.~\eqref{6.8}, that the solutions with $n=0$ exist only if
\begin{equation}
p_{z}=-\zeta\mathrm{sign}\left(g^{0}\right)\left|p_{z}\right|\,.\label{6.9}
\end{equation}
In this case, one gets that
\begin{align}
\,^{\zeta}\psi_{0,s}(\rho) & =C_{0}I_{0,s}(\rho)\begin{pmatrix}1\\
0
\end{pmatrix},\quad\begin{cases}
p_{z}<0, & \zeta=+1,\\
p_{z}>0, & \zeta=-1,
\end{cases}\quad g^{0}>0,\quad J_{z}>0\,,\nonumber \\
\,^{\zeta}\psi_{0,s}(\rho) & =C_{0}I_{0,s}(\rho)\begin{pmatrix}0\\
1
\end{pmatrix},\quad\begin{cases}
p_{z}>0, & \zeta=+1,\\
p_{z}<0, & \zeta=-1,
\end{cases}\quad g^{0}<0,\quad J_{z}<0\,.\nonumber \\
I_{0,s}(\rho) & =\frac{\left(-1\right)^{s}}{\sqrt{s!}}\rho^{s/2}e^{-\rho/2},\quad C_{0}=\sqrt{2\omega\left|g^{0}\right|}\,.\label{6.10}
\end{align}

Note that it is also possible to construct the set of solutions of the Dirac equation using the set in Eq.~(\ref{4.19g}). For $\chi=\mathrm{sign}(g^{0})$, we obtain the following: 
\begin{align}
 & I_{N+\left|\alpha\right|,N}\left(\rho\right)=I_{n,s}\left(\rho\right)\begin{cases}
1 & \ell\geq0\\
\left(-1\right)^{\ell} & \ell<0
\end{cases},\quad n=N_{J_{z},\chi=1}\in\mathbb{Z},\quad s=n-\ell\in\mathbb{Z},\quad g^{0}>0\,,\nonumber \\
 & I_{N+\left|\alpha\right|,N}\left(\rho\right)=I_{n,s}\left(\rho\right)\begin{cases}
\left(-1\right)^{\ell-1} & \ell\geq1\\
1 & \ell<1
\end{cases},\quad n=N_{J_{z},\chi=-1}\in\mathbb{Z},\quad s=n+\ell-1\in\mathbb{Z},\quad g^{0}<0\,,\label{7.10}
\end{align}
where the spectrum is defined by Eq.~(\ref{4.22-2}), and the parameter $s$ results from Eq.~(\ref{4.17aa}). Moreover, these parameters are positive integers, $n,s\in\mathbb{N}$, and condition in Eq.~(\ref{4.19f}) is satisfied automatically. Thus, we arrive at the same set of solutions in Eq.~(\ref{6.8}) since the phase factors $(-1)^{\ell}$ and $(-1)^{\ell-1}$ can be discarded without loss of generality.

\subsection{Calculating the current along the axis of rotation\label{sec:CURRCALCM0}}

Let us calculate the average hydrodynamic current of neutrinos and antineutrinos in a rotating frame of reference in local coordinates $(x^{\mu})$ along the axis of rotation,
\begin{equation}
J_{\zeta}^{3}=\sum_{n,s=0}^{\infty}\int dp_{z}\left[\overline{\,^{\zeta}\psi_{w}(x)}\gamma^{3}(x)\,^{\zeta}\psi_{w}(x)\rho_{\zeta}\left(\,^{\zeta}E_{n}\zeta\right)\right]\,,\label{7.1}
\end{equation}
where the measure of integration $dp_{z}$ follows from Eq.~(\ref{ort1}), $\rho_{\zeta}(E)=\{\exp(\beta[E-\zeta\mu_{A}])+1\}^{-1}$ is the Fermi-Dirac distribution for active neutrinos, $\beta=1/T$ is the inverse temperature, $\mu_{A}$ is the chemical potential of active neutrinos, $\,^{\zeta}E_{n}\zeta=\mathscr{E}_{n}+\zeta(g^{0}-\omega J_{z})$.

Using the complete set of solutions in Eq.~(\ref{6.8}), we obtain that
\begin{align}
 & \overline{\,^{\zeta}\psi_{w}(x)}\gamma^{3}(x)\,^{\zeta}\psi_{w}(x)=\zeta\mathrm{sign}\left(p_{z}\right)\frac{1}{2\pi}\left\{ \left(1-\delta_{n,0}\delta_{\zeta,\mathrm{sign}\left(g^{0}p_{z}\right)}\right)\varphi_{n',s}^{2}(\rho)\right.\nonumber \\
 & \times\left.\left[-\frac{\omega\left|g^{0}\right|}{\left|p_{z}\right|^{2}}n+3\left(\frac{\omega\left|g^{0}\right|}{\left|p_{z}\right|^{2}}\right)^{2}n^{2}+\dots\right]\left(\varphi_{n,s}^{2}(\rho)+\varphi_{n-1,s}^{2}(\rho)\right)\right\} \,,\label{7.2}\\
 & n'=n-\frac{1+\zeta\mathrm{sign}\left(g^{0}p_{z}\right)}{2}\,.\nonumber 
\end{align}
Note that $\,^{\zeta}E_{n}\zeta=\mathscr{E}_{n}+\zeta(g^{0}-\omega J_{z})$ is an even function of $p_{z}$. Thus, when integrating over $p_{z}$ in Eq.~(\ref{7.1}) from $-\infty$ to $+\infty$, only the first term in Eq.~(\ref{7.2}) will make a nontrivial contribution to Eq.~(\ref{7.1}):
\begin{equation}
J_{\zeta}^{3}=\zeta\frac{g^{0}\omega}{\pi}\int dp_{z}\left(1-\delta_{n,0}\delta_{\zeta,\mathrm{sign}\left(g^{0}p_{z}\right)}\right)\sum_{s=0}^{\infty}\rho_{\zeta}\left(\,^{\zeta}E_{0}\zeta\right)I_{0,s}^{2}\left(\rho\right)\,.\label{7.3}
\end{equation}
Taking into account Eq.~(\ref{6.10}) we have that
\begin{align}
 & J_{\zeta}^{3}=-\frac{g^{0}\omega}{\pi}\exp\left(-\rho\right)\sum_{s=0}^{\infty}\frac{\rho^{s}}{s!}\int_{0}^{\infty}\rho_{\zeta}\left(\,^{\zeta}E_{0}\zeta\right)dp_{z}\,,\nonumber \\
 & \,^{\zeta}E_{0}\zeta=\left|p_{z}\right|+\zeta\left[g^{0}-\mathrm{sign}\left(g^{0}\right)\omega\left(s+\frac{1}{2}\right)\right]\,.\label{7.4}
\end{align}
We integrate over $p_{z}$ in Eq.~(\ref{7.4}) and obtain:
\begin{align}
J_{\zeta}^{3} & =-\frac{g^{0}\omega}{\pi\beta}\exp\left(-\rho\right)\sum_{s=0}^{\infty}\frac{\rho^{s}}{s!}\ln\left[1+\exp\left(A+B_{s}\omega\right)\right]\, ,\label{7.5}\\
A & =\zeta\beta\left(\mu_{A}-g^{0}\right),\quad B_{s}=\zeta\beta\mathrm{sign}\left(g^{0}\right)\left(s+\frac{1}{2}\right)\,.
\end{align}
The series in $s$ in Eq.~(\ref{7.5}) converges for all $\rho$, which can easily be shown by applying d'Alembert's convergence test to this expression. 

It follows from Eq.~(\ref{7.5}) that
\begin{align}
J_{\zeta=1}^{3}-J_{\zeta=-1}^{3} & =-\frac{g^{0}\omega}{\pi}\exp\left(-\rho\right)\sum_{s=0}^{\infty}\frac{\rho^{s}}{s!}\left[\mu_{A}-g^{0}+\mathrm{sign}\left(g^{0}\right)\omega\left(s+\frac{1}{2}\right)\right]\label{7.6}\\
 & =-\frac{g^{0}\omega}{\pi}\left[\mu_{A}-g^{0}+\mathrm{sign}\left(g^{0}\right)\left(\frac{\omega}{2}+\left|g^{0}\right|\omega^{2}r^{2}\right)\right]\,.\nonumber 
\end{align}
The series in Eq.~(\ref{7.5}) can be calculated in the approximation of small $\omega$ by expanding the logarithm over the small parameter $B_{s}\omega$:
\begin{align}
 & \ln\left[1+\exp\left(A+B_{s}\omega\right)\right]=\ln\left(1+e^{A}\right)+\frac{e^{A}}{1+e^{A}}B_{s}\omega+\frac{\left(B_{s}\omega\right)^{2}}{4\left(1+\cosh A\right)}\nonumber \\
 & -\frac{\sinh^{4}\left(\frac{A}{2}\right)}{3\sinh^{3}A}\left(B_{s}\omega\right)^{3}+\frac{\cosh A-2}{192\cosh^{4}\left(\frac{A}{2}\right)}\left(B_{s}\omega\right)^{4}+O\left(\omega^{5}\right)\,.\label{7.7}
\end{align}
Substituting Eq.~(\ref{7.7}) into Eq.~(\ref{7.5}) and summing over $s$, we obtain that
\begin{align}
J_{\zeta=1}^{3} & =-\frac{g^{0}\omega}{\pi\beta}\left\{ \ln\left[1+e^{\beta\left[\mu_{A}+\mathrm{sign}\left(g^{0}\right)\omega/2-g^{0}\left(1-\omega^{2}r^{2}\right)\right]}\right]\right.\nonumber \\
 & +\left.\frac{\beta^{2}\omega^{2}}{8}\frac{1-\mathrm{\frac{5}{6}\beta\omega\,sign}\left(g^{0}\right)\tanh\left[\frac{\beta\left(\mu_{A}-g^{0}\right)}{2}\right]}{\cosh^{2}\left[\frac{\beta\left(\mu_{A}-g^{0}\right)}{2}\right]}\left|g^{0}\right|\omega r^{2}\right\} +O\left(\omega^{6}\right)\,.\label{7.8}
\end{align}
The expression for $J_{\zeta=-1}^{3}$ follows directly from Eqs.~(\ref{7.6}) and~(\ref{7.8}). 

Now, we consider the situation of electron neutrinos propagating in dense matter of a neutron star. In this situation, one gets that $g^{0}=-G_\mathrm{F}n_{n}/\sqrt{2}$ (see Ref.~\cite{Dvo15}). Then, we keep the leading term in $\omega$ in Eq.~\eqref{7.8}, as well as in the analogous expression for $J_{\zeta=-1}^{3}$. Eventually, we get that
\begin{align}
  & J_{\zeta=1}^{3}+J_{\zeta=-1}^{3}=G_{F}n_{n}\frac{\omega T}{\pi\sqrt{2}}F(y)+O\left(\omega^{2}\right),
  \quad
  F(y)=2\ln\left(1+e^{y}\right)-y\,,\label{7.9}\\
  & y=\frac{\mu_{A}-g^{0}}{T}\,.\nonumber 
\end{align}
which reproduces the result in Ref.~\cite{Dvo15}.

The expression for the current in Eq.~\eqref{7.8} contains both the leading term, linear in $\omega$, and nonlinear contributions. These additional terms result from the exact accounting for the noninertial effects and the neutrino matter interaction in solving of the Dirac equation in Sec.~\ref{sec:DIRACEQSOL}. It is the advantage of the present work with respect to Ref.~\cite{Dvo15}.

Let us estimate the typical recoil velocity $v$, acquired by a neutron star by the asymmetric neutrino emission given by Eq.~\eqref{7.9}. One has that $v = P/M$, where $P$ is the total momentum carried away by neutrinos and $M$ is the neutron star mass. We estimate $P$ as $P \sim J \bar{E} S \Delta t$, where $\bar{E}$ is the mean neutrino energy, $S = \pi R^2$ is the equatorial cross section of a star, $R$ is the stellar radius, and $\Delta t$ is the time of the neutrino emission. Finally, using Eq.~\eqref{7.9}, we get that
\begin{equation}\label{eq:recvel}
  v = \frac{3\bar{E}G_\mathrm{F}\omega T F(y)}{4\pi\sqrt{2}R m_n}\Delta t,
\end{equation}
where $m_n$ is the neutron mass.

We assume that $\bar{E} \sim T \sim \mu_A \sim 10\,\text{MeV}$, $R \sim 10\,\text{km}$, $\omega \sim 10^3\,\text{s}^{-1}$, and $\Delta t \sim 10^9\,\text{yr}$, i.e. we consider a quite old neutron star. In this case, $y \approx 1$ and $F(1) \approx 1.6$. Based on Eq.~\eqref{eq:recvel}, we obtain that $v \approx 6 \,\text{cm}\cdot \text{s}^{-1}$. One can see that the electroweak contribution to the linear pulsar velocity is quite small.

\section{Neutrino oscillations in slowly rotating matter}

In this section, we study neutrino flavor oscillations in rotating matter. Neutrino oscillations are known to happen only for massive particles. Hence, unlike Sec.~\ref{sec:INDCURR}, where we considered massless neutrinos, we take into account the particle masses here.

\subsection{Dirac equation for massive neutrinos in slowly rotating matter\label{sec:MASSSLOWROT}}

The solution of the Dirac equation for a massive neutrino interacting with matter in the noninertial frame is possible if we assume that the angular velocity is small. Luckily, it is the case for the majority of realistic neutrinos. Indeed, if we consider the neutrino propagation inside a millisecond pulsar having the rotation period $\tau = 10^{-3}\,\text{s}$ and the radius $R = 10\,\text{km}$, the linear velocity on the stellar equator $v=2\pi/\tau = 0.2 \ll 1$. It is the maximal possible matter velocity for astrophysical neutrinos. In practice, the contributions of the matter vorticity is smaller.

Analogously to Sec.~\ref{sec:DIRACEQSOL}, here, we consider a single neutrino mass eigenstate. The mixing between different neutrino types is accounted for shortly in Sec.~\ref{sec:NUOSC}.

In the limit of small $\omega$, we look for solutions of Dirac Eq.~(\ref{3.5}) in the form,
\begin{align}
 & \psi_{w}(r)=S\psi(r),\quad S=\exp\left\{ -\frac{1}{4}\ln\left[\frac{1-\omega^{2}r^{2}}{\left(1+\omega r\right)^{2}}\right]\gamma^{02}\right\} \,,\quad S^{\dag}=S^{\mathrm{T}}\,,\nonumber \\
 & \lim_{\omega r\rightarrow\infty}S=\frac{1-i\gamma^{02}}{\sqrt{2}},\quad S=e^{\gamma^{02}\omega r/2},\quad\omega\rightarrow0\,.\label{8.1}
\end{align}
Then, we obtain the equation for the function $\psi(r)$
\begin{align}
\left\{ i\gamma^{1}\left(\frac{d}{dr}+\frac{1}{2r}\right)+\frac{E}{\sqrt{1-\omega^{2}r^{2}}}\left(\gamma^{0}+\omega r\gamma^{2}\right)-J_{z}\frac{\sqrt{1-\omega^{2}r^{2}}}{r}\gamma^{2}\right.\nonumber \\
\left.-p_{z}\gamma^{3}-\frac{\omega}{2\left(1-\omega^{2}r^{2}\right)}\gamma^{3}\gamma^{5}-\frac{1}{2}g^{0}\sqrt{1-\omega^{2}r^{2}}\gamma^{0}\left(1-\gamma^{5}\right)-m\right\} \psi(r)=0\,.
\label{8.2}
\end{align}
We will assume that the rotation of matter is slow and neglect the terms $\sim(\omega r)^{2}$ at $\omega r\rightarrow0$ in Eq.~(\ref{8.2}):
\begin{align}
\mathscr{D}_{\omega}(r)\psi(r)=0,\quad & \mathscr{D}_{\omega}(r)=\left[i\gamma^{1}\left(\frac{d}{dr}+\frac{1}{2r}\right)-\gamma^{2}\left(\frac{J_{z}}{r}-\omega rE\right)+\gamma^{0}\left(E-\frac{g^{0}}{2}\right)-\gamma^{3}p_{z}\right.\nonumber \\
 & \qquad\,\,\,\left.+\frac{g^{0}}{2}\gamma^{0}\gamma^{5}-\frac{\omega}{2}\gamma^{3}\gamma^{5}-m\right]\,.\label{8.2b}
\end{align}

Note that Eq.~(\ref{8.1}) is the Dirac equation for the following tetrad
\begin{align}
\tilde{e}_{0}^{.\mu}(x) & =\frac{1}{\sqrt{1-\omega^{2}r^{2}}}\delta^{\mu0},\quad\tilde{e}_{1}^{.\mu}(x)=\delta^{\mu2}\,,\nonumber \\
\tilde{e}_{2}^{.\mu}(x) & =\frac{\omega r}{\sqrt{1-\omega^{2}r^{2}}}\delta^{\mu0}+\frac{\sqrt{1-\omega^{2}r^{2}}}{r}\delta^{\mu2},\qquad\tilde{e}_{3}^{.\mu}(x)=\delta^{\mu3}\,.\label{8.3}
\end{align}
The non-trivial fact is the existence of the second-order symmetry operator for Eq.~(\ref{8.2b}):
\begin{align}
\hat{M} & =\frac{d^{2}}{dr^{2}}+\frac{1}{r}\frac{d}{dr}+i\left(\frac{J_{z}}{r^{2}}+\omega E\right)\gamma^{12}\nonumber \\
 & -\left(J_{z}^{2}+\frac{1}{4}\right)\left(\omega^{2}+\frac{1}{r^{2}}\right)-\left(\omega Er\right)^{2}+\left(E-\omega J_{z}\right)^{2}-g^{0}\left(E-\frac{g^{0}}{2}\right)-p_{z}^{2}-m^{2}\,.\label{8.5}
\end{align}
This fact is true only for slow rotation and will allow us to find a complete set of solutions of Eq.~(\ref{8.2b}). 

We look for solutions of Eq.~(\ref{8.2b}) in the class of functions that are eigenfunctions of the integral of motion in Eq.~(\ref{8.5}),
\begin{equation}
\hat{M}\psi_{\kappa}(r)=\kappa\psi_{\kappa}(r)\,.\label{8.6}
\end{equation}
Using Eq.~(\ref{8.2b}), we express the derivative $\psi'_{\kappa}(r)$ and substitute it into in Eq.~(\ref{8.6}). As a result, we obtain the algebraic equation
\begin{equation}
\Omega\psi_{\kappa}(r)=\kappa\psi_{\kappa}(r),\quad\Omega=\left\{ \left[\omega p_{z}-g^{0}\left(E-\frac{g^{0}}{2}\right)\right]-m\left(\omega\gamma^{03}+g^{0}\gamma^{0}\gamma^{5}\right)\right\} \,.\label{8.7}
\end{equation}
Based on the fact that $\det\Omega=0$, one gets the compatibility condition,
\begin{equation}
\kappa^{2}+m^{2}\left[\left(g^{0}\right)^{2}-\omega^{2}\right]=\left\{ g^{0}E-\left[\omega p_{z}+\frac{\left(g^{0}\right)^{2}}{2}\right]\right\} ^{2}\,.\label{8.8}
\end{equation}
We can rewrite Eq.~\eqref{8.8} in the form,
\begin{equation}
E=\frac{g^{0}}{2}+\frac{\omega p_{z}}{g^{0}}+\zeta\sqrt{\frac{\kappa^{2}+m^{2}\left[\left(g^{0}\right)^{2}-\omega^{2}\right]}{\left(g^{0}\right)^{2}}},\quad\zeta=\pm1\,.\label{8.8a}
\end{equation}
The algebraic Eq.~(\ref{8.7}) implies that the solution of Eqs.~(\ref{8.2b})-(\ref{8.6}) can be represented as a linear combination,
\begin{equation}
\psi_{\kappa}(r)=\varphi_{-}(r)U_{-}+\varphi_{+}(r)U_{+},\quad U_{\pm}^{\dagger}U_{\pm}=1\,,\label{8.8b}
\end{equation}
where $\varphi_{\pm}(r)$ are some functions, and $U_{\pm}$ are orthonormal eigenvectors of the matrix $\Omega$,
\begin{align}
U_{-} & =\frac{1}{\sqrt{1+C_{-}^{2}}}\begin{pmatrix}A_{-}\\
0\\
1\\
0
\end{pmatrix},\quad U_{+}=\frac{1}{\sqrt{1+C_{+}^{2}}}\begin{pmatrix}0\\
A_{+}\\
0\\
1
\end{pmatrix},\quad C_{\pm}=\frac{\kappa-\omega p_{z}-g^{0}\left(E-\frac{g^{0}}{2}\right)}{m\left(g^{0}\pm\omega\right)}\,,\nonumber \\
 & A_{\pm}=\frac{\kappa+p_{z}\omega-g^{0}\left(E-g^{0}/2\right)}{m\left(g^{0}\pm\omega\right)}\,.\label{8.9}
\end{align}

Substituting Eq.~(\ref{8.8b}) into Eq.~(\ref{8.6}), we obtain a linear differential equation of the following form for the functions $\varphi_{\pm}(r)$:
\begin{align}
 & \left\{ \frac{d^{2}}{dr^{2}}+\frac{1}{r}\frac{d}{dr}-\frac{\left(J_{z}-\chi/2\right)^{2}}{r^{2}}+E^{2}\left(1-\omega^{2}r^{2}\right)\right.\nonumber \\
 & \left.+\frac{\omega E}{2}\left(J_{z}+\chi/2\right)-\frac{\omega^{2}}{4}-g^{0}\left(E-\frac{g^{0}}{2}\right)-p_{z}^{2}-m^{2}\right\} \varphi_{\chi}(r)=\kappa\varphi_{\chi}(r),\quad\chi=\pm1\,.\label{8.10}
\end{align}

As a result of substituting Eq.~(\ref{8.8b}) into Eq.~(\ref{8.2b}), we obtain two sets of solutions of the Dirac Eq.~(\ref{8.2b}):
\begin{align}
 & \psi_{\chi}(r)=\varphi_{\chi}(r)U_{\chi}+\varrho_{\chi}(r)U_{-\chi}\,,\nonumber \\
 & \varrho_{\chi}(r)=\frac{2i\left(g^{0}+\chi\omega\right)}{B_{\chi}}\sqrt{\frac{1+C_{\chi}^{2}}{1+C_{-\chi}^{2}}}\left[\varphi'_{\chi}(r)-\chi\left(\frac{J_{z}-\chi/2}{r}-\omega Er\right)\varphi_{\chi}(r)\right],\quad\chi=\pm\,,\nonumber \\
 & B_{\chi}=2\kappa+\chi\omega\left(2E+\chi\omega\right)-g^{0}\left(g^{0}+\chi\omega+2\chi p_{z}\right)\,,\label{8.11}
\end{align}
where the functions $\varphi_{\chi}(r)$ are solutions of Eq.~(\ref{8.10}). Solutions with different $\chi$ are linearly dependent. To derive Eq.~\eqref{8.11}, we take into account Eq.~(\ref{8.10})

By replacing $\rho=\omega|E|r^{2}$, Eq.~(\ref{8.10}) becomes identical to Eq.~(\ref{4.16}) with Laguerre polynomials, in which
\begin{equation}
\lambda=-\frac{1}{4\left|E\right|\omega}\left[\kappa-E^{2}-2\omega E\left(J_{z}+\frac{\chi}{2}\right)+\frac{\omega^{2}}{4}+g^{0}\left(E-\frac{g^{0}}{2}\right)+p_{z}^{2}+m^{2}\right]\,.\label{8.12}
\end{equation}

Bounded solutions of Eq.~(\ref{8.10}) have the form,
\begin{align}
 & \varphi_{\chi}(r)=\varphi_{n,s}(r)=\sqrt{2\left|E\right|\omega}I_{n,s}\left(\omega\left|E\right|r^{2}\right),\quad n,s\in\mathbb{Z}_{+}\,,\label{8.13}\\
 & \lambda=\lambda_{n}=n+\frac{1-\mathrm{sign}\left(E\right)\alpha}{2}\,,\\
 & J_{z}=\mathrm{sign}(E)\left(\frac{1}{2}-s-n_{\chi}\right),\quad n_{\chi}:=n+\frac{1-\mathrm{sign}\left(\chi E\right)}{2}\,,\label{8.13b}
\end{align}
where, from the requirement the $\lambda=\lambda_{n}$, one gets the expression for $\kappa$:
\begin{equation}
\kappa=-4\omega\left|E\right|n_{\chi}+\left(E-\frac{g^{0}}{2}\right)^{2}+\frac{\left(g^{0}\right)^{2}-\omega^{2}}{4}-m^{2}-p_{z}^{2}\,.\label{8.14}
\end{equation}

Substituting Eq.~(\ref{8.14}) into Eq.~(\ref{8.8}), we obtain the algebraic equation for the determination of the energy spectrum,
\begin{align}
 & \left\{ g^{0}E-\left[\omega p_{z}+\frac{\left(g^{0}\right)^{2}}{2}\right]\right\} ^{2}\nonumber \\
 & -\left[4\omega\left|E\right|n_{\chi}-\left(E-\frac{g^{0}}{2}\right)^{2}-\frac{\left(g^{0}\right)^{2}-\omega^{2}}{4}+m^{2}+p_{z}^{2}\right]^{2}=m^{2}\left[\left(g^{0}\right)^{2}-\omega^{2}\right]\,.\label{8.15}
\end{align}
If we set the mass equal to zero, $m=0$, then we immediately obtain the following solutions:
\begin{align}
\left|E_{S}\right| & =2\omega n_{\chi}+\sqrt{\left(2\omega n_{\chi}\right)^{2}+\left(p_{z}+\frac{\omega}{2}\right)^{2}},\quad E_{S}=\zeta\left|E_{S}\right|,\quad\zeta=\pm\,,\nonumber \\
\left|E_{A}\right| & =\zeta g^{0}+2\omega n_{\chi}+\sqrt{\left(\zeta g^{0}+2\omega n_{\chi}\right)^{2}+\left(p_{z}-\frac{\omega}{2}\right)^{2}-\left(g^{0}\right)^{2}}\quad E_{A}=\zeta\left|E_{A}\right|\,,\label{8.2.1}
\end{align}
The result (\ref{8.2.1}) matches the results of Ref.~\cite{Dvo14}.

In case of a nonzero mass $m\neq 0$, under the approximation of small $\omega$, we obtain, based on Eq.~\eqref{8.15},
\begin{equation}
E=\frac{g^{0}-\varepsilon_{1}\omega}{2}+\zeta\left(2+\varepsilon_{1}\frac{g^{0}}{p_{z}}\right)\omega n_{\chi}+\varepsilon_{2}\sqrt{\left(p_{z}+\zeta\frac{g^{0}}{p_{z}}\omega n_{\chi}+\varepsilon_{1}\frac{g^{0}}{2}\right)^{2}-\left(\frac{g^{0}}{p_{z}}\omega n_{\chi}\right)^{2}}+O\left(\omega^{2}\right)\,,\label{8.2.2}
\end{equation}
where the signs of $\varepsilon_{1}=\pm$ and $\varepsilon_2=\pm$ are chosen such that $\zeta=\mathrm{sign}E$. Note that the approximate solution in Eq.~(\ref{8.2.2}) is singular in $p_{z}$ and becomes meaningless when $p_{z}=0$.

Using the results of Appendix~\ref{sec:ENLEV}, one gets four roots of Eq.~\eqref{8.15},
\begin{equation}
E=\frac{g^{0}}{2}+2\zeta n_{\chi}\omega+\frac{1}{2}\left(\varepsilon_{1}\sqrt{2\xi_{1}}+\varepsilon_{2}\sqrt{-2\left(\mathfrak{p}+\xi_{1}\right)-\varepsilon_{1}\mathfrak{q}\sqrt{\frac{2}{\xi_{1}}}}\right),
\quad
\varepsilon_{2}=\pm1\,,\label{8.2.11}
\end{equation}
where the signs of $\varepsilon_{1}=\pm$ and $\varepsilon_2=pm$ are chosen such that $\zeta=\mathrm{sign}E$. For convenience, we set $\chi=\mathrm{sign}(E)$ and $n_{\chi}=n$.

Then for $E<0$ the solution set (\ref{8.11}) takes the form
\begin{align}
 & \psi(r)=\mathcal{N}_{n,s}^{(-)}\left[\varphi_{n,s}(r)U_{-}+F_{n,s}^{(-)}\varphi_{n-1,s}(r)U_{+}\right]\,,\quad E<0\,,\nonumber \\
 & F_{n,s}^{(-)}=4i\sqrt{\frac{1+C_{+}^{2}}{1+C_{-}^{2}}}\frac{\left(g^{0}-\omega\right)\sqrt{\left|E\right|\omega n}}{2\kappa+\omega\left(2\left|E\right|+\omega\right)-g^{0}\left(g^{0}-\omega-2p_{z}\right)}\,,\nonumber \\
 & \mathcal{N}_{n,s}^{(-)}=\left\{ 1+\left|F_{n,s}^{(-)}\right|^{2}\right\} ^{-1/2}\,,\label{8.16}
\end{align}
and for $E>0$ it will be determined by the expression
\begin{align}
 & \psi(r)=\mathcal{N}_{n,s}^{(+)}\left[\varphi_{n,s}(r)U_{+}+F_{n,s}^{(+)}\varphi_{n-1,s}(r)U_{-}\right]\,,\quad E>0\,,\nonumber \\
 & F_{n,s}^{(+)}=-4i\sqrt{\frac{1+C_{-}^{2}}{1+C_{+}^{2}}}\frac{\left(g^{0}+\omega\right)\sqrt{\left|E\right|\omega n}}{2\kappa+\omega\left(2\left|E\right|+\omega\right)-g^{0}\left(g^{0}+\omega+2p_{z}\right)}\,,\nonumber \\
 & \mathcal{N}_{n,s}^{(+)}=\left\{ 1+\left|F_{n,s}^{(+)}\right|^{2}\right\} ^{-1/2}\,.\label{8.17}
\end{align}
In this case, states with minimum energy, which correspond to $n=0$, exist for any value of the quantum number $p_{z}$, in contrast to the massless case.

\subsubsection{Induced current of massive neutrinos\label{sec:CURRMASS}}

It is interesting to provide the expression for the induced current, which was studied in Sec.~\ref{sec:CURRCALCM0} for $m=0$, in the massive case. Based on the results of Sec.~\ref{sec:MASSSLOWROT}, one gets for small $\omega$ that 
\begin{equation}
\sum_{s=0}^{\infty}\overline{\,^{\zeta}\psi(x)}\gamma^{3}(x)\,^{\zeta}\psi(x)=\zeta\frac{\omega\left(\left|p_{z}\right|+g^{0}/2\right)}{\pi}\left[1-\zeta\frac{g^{0}/2}{\sqrt{\left(\left|p_{z}\right|+g^{0}/2\right)^{2}+m^{2}}}\right]+O\left(\omega^{2}\right)\,.\label{8.18}
\end{equation}
Then states with $n>0$ also contribute to the current:
\begin{align}
J_{\zeta}^{3} & =\sum_{n,s=0}^{\infty}\int dp_{z}\left[\overline{\,^{\zeta}\psi(x)}\gamma^{3}(x)\,^{\zeta}\psi(x)\rho_{\zeta}\left(\,^{\zeta}E_{n}\zeta\right)\right]\nonumber \\
 & =\frac{\zeta\omega}{\pi}\sum_{n=0}^{\infty}\int dp_{z}\left(\left|p_{z}\right|+g^{0}/2\right)\left[1-\zeta\frac{g^{0}/2}{\sqrt{\left(\left|p_{z}\right|+g^{0}/2\right)^{2}+m^{2}}}\right]\rho_{\zeta}\left(\,^{\zeta}E_{n}\zeta\right)+O\left(\omega^{2}\right)\,.\label{8.19}
\end{align}
Despite we do not provide the explicit expression for the current since, one can see that $J_{\zeta}^{3} \neq 0$ in Eq.~\eqref{8.19}. This fact can be illustrated as follows. Equation~(\ref{8.19}) can be approximately estimated by setting $g^0 = 0$, and we obtain a nonzero expression,
\begin{equation}\label{eq:Jmassive}
	J_{\zeta}^{3}|_{g^0 = 0} \sim \frac{\zeta \omega}{\pi\beta^2} 
	\left[ \coth\left(\omega\beta\right) - 1 \right]
	\exp\left[\beta\left(\frac{3}{2}\omega + \zeta \mu_A - m\right)\right]
	+O\left(\omega^{2}\right),\quad \beta\rightarrow 0\,.
\end{equation}

The current in Eq.~\eqref{eq:Jmassive} depends on the chemical potential of active neutrinos, which are left particles in the standard model. Thus, this current, in fact, should be attributed to left chiral projection of neutrino states. The axial currents generation in frames of the CVE for massive particles was studied, e.g., in Ref.~\cite{FlaFuk17}. We also mention that, in our case, the noninertial effects can also contribute to the magnitude of the current.


\subsection{Neutrino oscillations in rotating matter\label{sec:NUOSC}}

In this section, we apply the results of Sec.~\ref{sec:MASSSLOWROT} to describe neutrino flavor oscillations in rotating matter. We restrict ourselves to case of the small angular velocity. However, we exactly take into account the neutrino masses. The neutrino mixing and the interaction with matter in flat space-time were studied in Sec.~\ref{sec:NUMIXMATT}. Now, we generalize these concepts to a curved space-time.

We consider the evolution of a system of two types of neutrinos $\nu=(\nu_{\alpha},\nu_{\beta})$. The generalization of Eqs.~\eqref{eq:masstermgen} and~\eqref{eq:mattintgen} reads
\begin{equation}
	\mathcal{L}=
	\sum_{\lambda=\alpha,\beta}\overline{\nu_{\lambda}}i\gamma^{\mu}(x)\left(\nabla_{\mu}+\Gamma_{\mu}\right)\nu_{\lambda}-\sum_{\lambda,\lambda'=\alpha,\beta}\overline{\nu_{\lambda}}\left(m_{\lambda\lambda'}+f_{\lambda}^{\mu}\delta_{\lambda\lambda'}\right)\nu_{\lambda'}\,,\label{5.1}
\end{equation}
where $m_{\lambda,\lambda'}$ are $f_{\lambda}^{\mu}$ are defined in Sec.~\ref{sec:NUMIXMATT}.

The flavor neutrino eigenstates $\psi_{a}$, $a=1,2$, are introduced analogously to Eq.~\eqref{eq:massstates}. In the case when only two types of neutrinos are present in the system, the mixing matrix has the form
\begin{equation}
	\left(U_{\lambda a}\right)=\begin{pmatrix}\cos\theta & -\sin\theta\\
		\sin\theta & \cos\theta
	\end{pmatrix}\,,\label{5.3}
\end{equation}
where $\theta$ is the vacuum mixing angle. 

Next, we will assume that $\nu_\mu$ corresponds to the index $\alpha=1$, and $\nu_e$ corresponds to the index $\beta=2$.
The generalization of Eq.~\eqref{eq:Lagrmattmass} has the form,
\begin{equation}
	\mathcal{L}=
	\sum_{a=1,2}\overline{\psi_{a}}\left[i\gamma^{\mu}(x)\left(\nabla_{\mu}+\Gamma_{\mu}\right)-m_{a}\right]\psi_{a}-\frac{1}{2}
	\sum_{a,b=1,2} g_{ab}^{\mu}\overline{\psi_{a}}\left(1-\gamma^{5}\right)\psi_{b}\,,\label{5.4}
\end{equation}
where $g_{ab}^{\mu}$ is given in Eq.~\eqref{eq:mattpotmass}.
Choosing the corotating frame we get that only $g_{ab}^{0}$ is nonvanishing. 

One gets the wave equations for the mass eigenstates form the Lagrangian in Eq.~(\ref{8.3}),
\begin{equation}
	\left[i\gamma^{\mu}(x)\left(\nabla_{\mu}+\Gamma_{\mu}\right)-m_{a}-\frac{1}{2}g_{a}^{0}\gamma_{0}(x)\left(1-\gamma^{5}\right)\right]\psi_{a}(x)=\frac{1}{2}\gamma_{0}(x)\sum_{b\neq a}g_{ab}^{0}\left(1-\gamma^{5}\right)\psi_{b}(x)\,.\label{5.6}
\end{equation}
One can see in Eq.~\eqref{5.6} that the wave equations for different mass eigenstates are coupled.

Following Ref.~\cite{Dvo11}, we will look for the solutions of Eq.~\eqref{5.6} in the form,
\begin{equation}
	\psi_{a}(x)=\frac{1}{2\pi}\sum_{n,s=0}^{\infty}\int_{-\infty}^{+\infty}dp_{z}\,A_{n,s,p_{z}}^{(a)}(t)\psi_{n,s,p_{z}}^{(a)}(r)e^{ip_{z}z+iJ_{z}\phi}e^{-iE_{a}t}\,,\label{5.7}
\end{equation}
where $\psi_{n,s,p_{z}}^{(a)}(r)$ is the complete set of solutions of Eqs.~(\ref{8.16})-(\ref{8.17}), in which $m:=m_{a}$, $g^{0}:=g_{a}^{0}$.
Since we use the relativistic quantum mechanics to describe the evolution of neutrinos, the coefficients $A_{n,s,p_{z}}^{(a)}(t)$ are $c$-number functions of time.

We choose the initial conditions:
\begin{align}
	& \nu_{\mu}\left(\mathbf{r},t=0\right)=0\,,\nonumber \\
	& \nu_{e}\left(\mathbf{r},t=0\right)=\frac{1}{2\pi}\psi_{n^{(0)},s^{(0)},p_{z}^{(0)}}^{(a)}(r)e^{ip_{z}^{(0)}z+iJ_{z}^{(0)}\phi}\,.\label{5.13}
\end{align}
which correspond to a situation when one has only $\nu_e$ and no $\nu_\mu$ initially.

Since will not take into account neutrino-antineutrino transitions, one has that $E_{a}\geq0$. Then, for small $\omega$ we have
\begin{equation}
	E_{a}=\frac{g_{a}^{0}+\omega}{2}+\left(2-\frac{g_{a}^{0}}{p_{z}}\right)\omega n+\sqrt{\left(p_{z}+\frac{g_{a}^{0}}{p_{z}}\omega n-\frac{g_{a}^{0}}{2}\right)^{2}-\left(\frac{g_{a}^{0}}{p_{z}}\omega n\right)^{2}}+O\left(\omega^{2}\right)\,\label{5.8}
\end{equation}
Using the approximations,
\begin{equation}
	\sqrt{1-\left(\omega r\right)^{2}}=1+O\left(\omega r\right)^{2},\quad\left(\gamma^{0}+\gamma^{2}\omega r\right)^{2}=I_{4}+O\left(\omega^{2}\right)\,,\label{5.9}
\end{equation}
on the basis of Eq.~(\ref{5.6}) we get the Schr\"{o}dinger equation for the  effective wavefunction,
\begin{equation}
	\Phi^\mathrm{T} = \left(A^{(1)}_{n^{(0)},s^{(0)}(t),p_z^{(0)}}\exp(-iE_1 t),
                        	A^{(2)}_{n^{(0)},s^{(0)}(t),p_z^{(0)}}\exp(iE_2 t)\right)\,,
\end{equation}
in the form
\begin{equation}
	i\frac{d}{dt}\Phi = \begin{pmatrix}
							E_1 & V^{12}_{n^{(0)},s^{(0)}(t),p_z^{(0)}} \\
							V^{21}_{n^{(0)},s^{(0)}(t),p_z^{(0)}} & E_2
						\end{pmatrix}\Phi\,.
\end{equation}
where
\begin{align}
	V_{n,s,p_{z}}^{ab} & :=\frac{g_{ab}^{0}}{2}\langle\psi_{n,s,p_{z}}^{(a)}\mid\gamma_{0}\left(1-\gamma^{5}\right)\mid\psi_{n,s,p_{z}}^{(b)}\rangle\nonumber \\
	& =g_{ab}^{0}\frac{\sqrt{\left(k_{a}+g_{a}^{0}\right)\left(k_{b}+g_{b}^{0}\right)}\left(k_{a}-g_{a}^{0}+2p_{z}-2m_{a}\right)\left(k_{b}-g_{b}^{0}+2p_{z}+2m_{b}\right)}{8p_{z}^{2}\sqrt{4m_{a}^{2}+\left(2p_{z}-g_{a}^{0}\right)\left(k_{a}-g_{a}^{0}+2p_{z}\right)}\sqrt{4m_{b}^{2}+\left(2p_{z}-g_{b}^{0}\right)\left(k_{b}-g_{b}^{0}+2p_{z}\right)}}\omega n\nonumber\\
	&+O\left(\omega^{2}\right)\,,\label{5.11}\\
	k_{a} & :=\sqrt{4m_{a}^{2}+\left(g_{a}^{0}-2p_{z}\right)}\,.
\end{align}

Then, the transition probability for $\nu_e \to \nu_\mu$ oscillations has the form,
\begin{align}
	\mathcal{P}_{\nu_{e}\rightarrow\nu_{\mu}}(t) & =\left|\langle\nu_{2}\left(0\right)\mid\nu_{1}\left(t\right)\rangle\right|^{2}=P_{max}\sin^{2}\left(\varUpsilon t\right)\,,\nonumber \\
	P_{max} & =\frac{\left[\left(V_{n^{(0)},s^{(0)},p_{z}^{(0)}}^{12}\cos^{2}\theta-V_{n^{(0)},s^{(0)},p_{z}^{(0)}}^{21}\sin^{2}\theta\right)+\Xi\sin2\theta\right]^{2}}{\left|\Xi^{2}+V_{n^{(0)},s^{(0)},p_{z}^{(0)}}^{12}V_{n^{(0)},s^{(0)},p_{z}^{(0)}}^{21}\right|}\,.\label{5.14}
\end{align}

For small $\omega$, we obtain the amplitude of the transition probability in Eq.~\eqref{5.14} in the form,
\begin{align}
	& P_{max}=\sin^2(2\theta) + \frac{8\omega n \left(V_{n^{(0)},s^{(0)},p_{z}^{(0)}}^{12}\cos^{2}\theta-V_{n^{(0)},s^{(0)},p_{z}^{(0)}}^{21}\sin^{2}\theta\right)}{k_{1}^{(0)}-k_{2}^{(0)}+g_{1}^{0}-g_{2}^{0}}+O\left(\omega^{2}\right)\,.\label{5.15}
\end{align}

It is interesting to analyze the resonance condition for neutrino oscillations in rotating matter. The function $P_{\max}$ in Eq.~\eqref{5.15} reaches the unit value when the condition
\begin{equation}
	\cos^2(2\theta) - \frac{2\omega n}{k_{1}^{(0)}-k_{2}^{(0)}+g_{1}^{0}-g_{2}^{0}}\left(V_{n^{(0)},s^{(0)},p_{z}^{(0)}}^{12} \cos^2\theta - V_{n^{(0)},s^{(0)},p_{z}^{(0)}}^{21} \sin^2\theta\right)\sin(2\theta)=0\,.\label{5.16}
\end{equation}
is fulfilled.

Equations~\eqref{5.14}-\eqref{5.16} generalize the expressions for neutrino oscillations in matter. In particular, the resonance condition in Eq.~\eqref{5.16} is the analogue of the MSW effect~\cite{Wol78,MikSmi85} in rotating matter accounting for the noninertial effects.

\section{Conclusion}\label{sec:CONCL}

In conclusion, we mention that we have analyzed the evolution of neutrinos in rotating matter on the basis of the exact solution of the Dirac equation in the noninertial frame. We have obtained this solution in two particular situations. 

First, in Sec.~\ref{sec:DIRACEQSOL}, we have analyzed the case of massless neutrinos in matter rotating with arbitrary angular velocity using the new approach of squaring the Dirac equation. It follows from Eq.~(\ref{7.10}) that the obtained set of solutions can also be described in terms of the Laguerre functions given in Eq.~(\ref{4.19g}). It suggests that the spectrum of massless neutrinos in rotating matter, which was obtained in Ref.~\cite{Dvo14}, is reduced to the expression for the neutrino spectrum obtained in Ref.~\cite{Olivera25}.

In Sec.~\ref{sec:MASSSLOWROT}, we have accounted for a nonzero neutrino mass in a slowly rotating matter taking into account the non-trivial symmetry of the Dirac equation, which corresponds to the second-order symmetry operator (the use of symmetry operators to obtain solutions to the Dirac equation is discussed in Refs. \cite{BrSh16,BagOb92,BagGit90book,BagGit14book,BagGitSmir04}).

Using the solutions obtained, we considered several applications important for astrophysical neutrinos. In Sec.~\ref{sec:INDCURR}, we have calculated the hydrodynamic current of neutrinos along the matter vorticity. This phenomenon is analogous to the CVE~\cite{Kha16}. We have obtained the nonzero electroweak contribution to this current in case of massless neutrinos. This our result generalizes the finding of Ref.~\cite{Dvo18b} where the current was derived in the flat spacetime. In the present work, we have obtained the expression for the current in the noninertial frame. We have reproduced the results of Ref.~\cite{Dvo18b} in case of a slowly rotating matter; cf. Eqs.~\eqref{7.7} and~\eqref{7.8}.

In Sec.~\ref{sec:CURRMASS}, we have calculated the neutrino current for massive particles. This current turns out to be nonzero for massive neutrinos. This our result has been obtained in the case of a slowly rotating matter. The CVE for massive particles was studied previously, e.g., in Ref.~\cite{FlaFuk17}. However, in our opinion, there is a nontrivial contribution to the current owing to noninertial effects.

Our result that one has a nonzero current of neutrinos along the rotation axis could have an implication for the explanation of great linear velocities of some pulsars. However, the numerical estimates provided in Sec.~\ref{sec:CURRCALCM0} point out that the electroweak contribution to the pulsar recoil velocity is negligible compared to other mechanisms.

The next application of the obtained solution of the Dirac equation covers neutrino flavor oscillations in rotating matter. Since neutrino oscillations can happen only for massive particles, we had to use the solution corresponding to massive neutrinos in a slowly rotating matter. In Sec.~\ref{sec:NUOSC}, we have derived that transition probability for $\nu_e \to \nu_\mu$ oscillations. We have obtained that the analogue of the MSW effect takes place in rotating matter. We have have derived the contribution of the matter vorticity to the resonance condition.


\appendix

\section{Properties of spinors in a rotating frame\label{sec:ROTFRAME}}

In this Appendix, we represent some differential geometry objects and the components of the Dirac equation in the noninertial rotating frame.

Based on Eq.~\eqref{1.3}, we rewrite the metric tensor $\left(g_{\mu\nu}(x)\right)$ ($\mu,\nu=0,\dots,3)$ and the reciprocal, as
\begin{align}
\left(g_{\mu\nu}(x)\right)= & \begin{pmatrix}1-\omega^{2}r^{2} & 0 & -\omega r^{2} & 0\\
0 & -1 & 0 & 0\\
-\omega r^{2} & 0 & -r^{2} & 0\\
0 & 0 & 0 & -1
\end{pmatrix}\,,\label{1.4}\\
\left(g^{\mu\nu}(x)\right)=\left(g_{\mu\nu}(x)\right)^{-1}= & \begin{pmatrix}1 & 0 & -\omega & 0\\
0 & -1 & 0 & 0\\
-\omega & 0 & \omega^{2}-\frac{1}{r^{2}} & 0\\
0 & 0 & 0 & -1
\end{pmatrix}\,.\nonumber 
\end{align}

The Christoffel symbols of a symmetric connection compatible with
the metric in Eq.~(\ref{1.3}) are defined using the metric tensor in Eq.~(\ref{1.4}):
\begin{equation}
\Gamma_{\nu\mu}^{\rho}(g)=\frac{1}{2}g^{\rho\tau}(x)\left[\partial_{\nu}g_{\tau\mu}(x)+\partial_{\mu}g_{\mu\tau}(x)-\partial_{\tau}g_{\nu\mu}(x)\right],\quad\partial_{\mu}:=\frac{\partial}{\partial x^{\mu}}\,.\label{1.4b}
\end{equation}
The non-zero components of the Christoffel symbols of the metric in Eq.~(\ref{1.3})
have the form,
\begin{equation}
\Gamma_{22}^{1}=-r,\quad\Gamma_{02}^{1}=\omega\Gamma_{22}^{1},\quad\Gamma_{00}^{1}=\omega^{2}\Gamma_{22}^{1},\quad\Gamma_{12}^{2}=\frac{1}{r},\quad\Gamma_{01}^{2}=\omega\Gamma_{12}^{2}\,.\label{1.4c}
\end{equation}

The metric tensor in Eq.~(\ref{1.4}) can be diagonalized, $\eta_{ab}=g_{\mu\nu}(x)e_{a}^{.\mu}(x)e_{b}^{.\nu}(x)$,
using the tetrad ($a=1,\dots,4)$:
\begin{align}
e_{0}^{.\mu}(x) & =\delta^{\mu0}-\omega\delta^{\mu2},\quad e_{1}^{.\mu}(x)=\delta^{\mu1}\,,\label{1.5}\\
e_{2}^{.\mu}(x) & =\frac{1}{r}\delta^{\mu2},\quad e_{3}^{.\mu}(x)=\delta^{\mu3}\,,\nonumber 
\end{align}
where $\delta^{\mu\nu}$ is the Kronecker delta and $\mathrm{det}\left(e_{a}^{.\mu}(x)\right)>0$.
Let us introduce the inverse tetrad $\left(e_{\mu}^{.a}(x)\right)=\left(e_{a}^{.\mu}(x)\right)^{-1}$,
which obeys $e_{a}^{.\mu}(x)e_{\mu}^{.b}(x)=\delta_{a}^{.b}$, $e_{\mu}^{.a}(x)e_{a}^{.\nu}(x)=\delta_{\mu}^{.\nu}$, as
\begin{align}
e_{0}^{.a}(x) & =\delta^{a0}+\omega r\delta^{a2},\quad e_{1}^{.a}(x)=\delta^{a1}\,,\label{1.6}\\
e_{2}^{.a}(x) & =r\delta^{a2},\quad e_{3}^{.a}(x)=\delta^{a3}\,.\nonumber 
\end{align}
We note that $g(x)=\mathrm{det}\left(g_{\mu\nu}(x)\right)=-\mathrm{det}\left(e_{a}^{.\mu}(x)\right)^{-2}=-r^{2}$.
The Levi-Civita tensor in the metric in Eq.~(\ref{1.3}) is defined by the
expression,
\[
e_{\mu\nu\sigma\rho}(x)=\sqrt{\left|g(x)\right|}\varepsilon_{\mu\nu\sigma\rho}=r\varepsilon_{\mu\nu\sigma\rho},\quad e^{\mu\nu\sigma\rho}(x)=\frac{\mathrm{sgn}\left(g(x)\right)}{\sqrt{\left|g(x)\right|}}\varepsilon^{\mu\nu\sigma\rho}=-\frac{1}{r}\varepsilon^{\mu\nu\sigma\rho},
\]
where $\varepsilon^{\mu\nu\sigma\rho}=\varepsilon_{\mu\nu\sigma\rho}$
is the Levi-Civita symbol, $\varepsilon_{0123}=+1$. 

The indices, numbering the tetrad, are raised and lowered using the
metric tensor $(\eta_{ab})$ and are denoted by Latin letters. The
Greek indices are raised and lowered using the metric tensor $\left(g_{\mu\nu}(x)\right)$.
In both cases, the indices take values from $0$ to $3$.

The Dirac matrices $\gamma^{\mu}(x)$ are introduced as an arbitrary
but fixed solution of the equation,
\begin{equation}
\left\{ \gamma^{\mu}(x),\gamma^{\nu}(x)\right\} :=\gamma^{\mu}(x)\gamma^{\nu}(x)+\gamma^{\nu}(x)\gamma^{\mu}(x)=2g^{\mu\nu}(x)I_{4}\,,\label{2.1}
\end{equation}
where $I_{N}$ is the identity matrix of size $N\times N$. We fix
the solution of Eq.~(\ref{2.1}) in the form $\gamma^{\mu}(x)=\gamma_{a}e_{\mu}^{.a}(x)$,
where $\gamma_{a}$ are the standard Dirac matrices which are coordinate independent. In the chiral
representation,
\begin{equation}
\gamma^{0}=\begin{bmatrix}0 & -I_{2}\\
-I_{2} & 0
\end{bmatrix},\quad\gamma^{k}=\begin{bmatrix}0 & \sigma_{k}\\
-\sigma_{k} & 0
\end{bmatrix},\quad k=1,\dots,3\,,\label{2.1b}
\end{equation}
where$\{\sigma_{1},\sigma_{2},\sigma_{3}\}$ are the Pauli matrices:
\begin{equation}
\sigma_{1}=\begin{pmatrix}0 & 1\\
1 & 0
\end{pmatrix},\quad\sigma_{2}=\begin{pmatrix}0 & -i\\
i & 0
\end{pmatrix},\quad\sigma_{3}=\begin{pmatrix}1 & 0\\
0 & -1
\end{pmatrix}\,.\label{4.5}
\end{equation}
Using Eqs.~\eqref{2.1}-\eqref{4.5}, one gets that
\begin{equation}
\gamma^{5}(x):=\frac{i}{4!}e_{\mu\nu\sigma\rho}(x)\gamma^{\mu}(x)\gamma^{\nu}(x)\gamma^{\sigma}(x)\gamma^{\rho}(x)=\gamma^{5}=i\gamma^{0}\gamma^{1}\gamma^{2}\gamma^{3}=\begin{bmatrix}I_{2} & 0\\
0 & -I_{2}
\end{bmatrix}\,.\label{2.2}
\end{equation}

The Fock-Ivanenko coefficients, or the spin connection, $\Gamma_{\mu}(x)$
are uniquely determined from the conditions 
\begin{equation}
[\gamma^{\mu}(x),\nabla_{\nu}+\Gamma_{\nu}(x)]=0,\quad\mathrm{Sp}\Gamma_{\mu}=0,\label{2.3}
\end{equation}
where $[A, B] = AB - BA$ is the commutator of operators $A$ and $B$. The explicit expression for the Fock-Ivanenko coefficients
results from Eq.~\eqref{2.3},
\begin{equation}
\Gamma_{\mu}(x)=-\frac{1}{4}\gamma_{\nu;\mu}(x)\gamma^{\nu}(x),\quad\gamma_{\nu;\mu}(x)=\gamma_{\nu,\mu}(g)-\Gamma_{\nu\mu}^{\tau}(x)\gamma_{\tau}(x)\,.\label{2.4}
\end{equation}
The action of the covariant derivative $\nabla_{\mu}$ on the tensor
is denoted by a semicolon, whereas the action of the ordinary derivative
is denoted by a comma, $\gamma_{\nu;\mu}(x):=\nabla_{\mu}\gamma_{\nu}(x)$ and
$\gamma_{\nu,\mu}(x):=\partial_{\mu}\gamma_{\mu}(x)$. 

In the metric in Eq.~(\ref{1.3}), the Fock-Ivanenko coefficients have the
form:
\begin{align}
\Gamma_{0}(x) & =\frac{\omega}{2}\gamma^{12},\quad\Gamma_{1}(x)=0\,,\nonumber \\
\Gamma_{2}(x) & =\frac{1}{\omega}\Gamma_{0}(x),\quad\Gamma_{3}(x)=0,\label{2.5}\\
\gamma^{ab} & =\frac{1}{2}[\gamma^{a},\gamma^{b}]:=\frac{1}{2}\left(\gamma^{a}\gamma^{b}-\gamma^{b}\gamma^{a}\right)\,.
\end{align}
The coefficients in Eq.~(\ref{2.5}) are nonzero, which corresponds to the
fact that the motion of a test particle in a non-inertial, in this
case, in the rotating frame of reference is equivalent to the interaction
of this particle with the gravitational field. 

\section{Properties of Laguerre functions\label{sec:LAGPROP}}

In this appendix, we list some useful properties Laguerre functions based on the Refs. \cite{BagGitSmir04,BagGit14book}.

In general, the Laguerre functions are defined by the relation,
\begin{equation}
I_{u,v}(\rho)=\frac{1}{\Gamma\left(u-v+1\right)}\sqrt{\frac{\Gamma\left(u+1\right)}{\Gamma\left(v+1\right)}}e^{-\rho/2}\rho^{(u-v)/2}\Phi\left(-v,u-v+1;\rho\right)\,,\label{5.5b}
\end{equation}
where $\Phi\left(\xi,\eta;\rho\right)$ is a degenerate hypergeometric function. If the parameter $v$ takes non-negative integer values, then the Laguerre functions in Eq.~(\ref{5.5b}) are related to the Laguerre polynomials $L_{n}^{\beta}(\rho)=(n!)^{-1}d^{n}(e^{-x}x^{n+\beta})/d\rho^{n}$,
\[
I_{u,n}(\rho)=\sqrt{\frac{n!}{\Gamma\left(u+1\right)}}e^{-\rho/2}\rho^{(u-n)/2}L_{n}^{u-n}(\rho),\quad n\in\mathbb{Z}_{+},\quad\mathbb{Z}_{+}=\left\{ 0,1,2,\dots\right\} \,.
\]

Laguerre functions in Eq.~(\ref{4.17b}) form a complete and orthonormal system,
\begin{eqnarray}
	& \int_{0}^{\infty}I_{s+\alpha,s}(\rho)I_{s'+\alpha,s'}(\rho)d\rho=\delta_{s,s'}\,,
	\label{4.20}\\
	&\sum_{n=0}^{\infty}I_{s+\alpha,s}(\rho)I_{s+\alpha,s}(\rho')=\delta_{\rho,\rho'}\,.
	\label{4.20b}
\end{eqnarray}
For large values of $\rho$, the Laguerre functions have the following asymptotics:
\begin{equation}
I_{n,s}(\rho)\sim\frac{(-1)^{s}}{\sqrt{n!s!}}\rho^{(m+s)/2}e^{-\rho/2},\quad\rho\rightarrow\infty\,.\label{4.19c}
\end{equation}
It may seem that the functions in Eq.~(\ref{4.17b}) for $n-s<0$ have a singularity at the point $\rho=0$. However, this is not the case if we keep in mind the property 
\[
L_{s}^{n-s}(\rho)=(-1)^{n-s}\frac{n!}{s!}\rho^{-(n-s)}L_{n}^{-(n-s)}(\rho)\,.
\]
Therefore, defining
\begin{equation}
I_{n,s}\left(\rho\right)=\begin{cases}
I_{n,s}\left(\rho\right), & n\geq s,\\
\left(-1\right)^{n-s}I_{s,n}\left(\rho\right), & n<s,
\end{cases}\label{4.19d}
\end{equation}
for small values of $\rho$. Taking into account Eq.~(\ref{4.19d}), we have
\begin{equation}
I_{n,s}(\rho)\sim\frac{\left(-1\right)^{n-s}}{\left|n-s\right|!}\left(\frac{n!}{s!}\right)^{\epsilon/2}\rho^{\left|n-s\right|/2},\quad\epsilon:=\mathrm{sign}\left(n-s\right),\quad\rho\rightarrow0\,.\label{4.19b}
\end{equation}
From Eqs.~(\ref{4.19c}) and~(\ref{4.19b}) it is easy to see that these functions tend to zero at $\rho\rightarrow0$ and $\rho\rightarrow\infty$.

\section{Energy spectrum for massive neutrinos in a slowly rotating matter\label{sec:ENLEV}}

In this Appendix, we show how to get the roots of Eq.~\eqref{8.15} to determine the energy spectrum of massive neutrinos in Sec.~\ref{sec:MASSSLOWROT}.

In general, Eq.~(\ref{8.15}) is an incomplete quartic equations
\begin{equation}
w^{4}+\mathfrak{p}w^{2}+\mathfrak{q}w+\mathfrak{r}=0,\quad w=E-\frac{g^{0}}{2}-2\zeta n_{\chi}\omega\,,\label{8.2.3}
\end{equation}
where $\zeta=\mathrm{sign}E$. The coefficients in Eq.~\eqref{8.2.3} are
\begin{align}
-\frac{\mathfrak{p}}{2} & =\left(2\omega n_{\chi}+\zeta\frac{g^{0}}{2}\right)^{2}+\frac{\omega^{2}}{4}+p_{z}^{2}+m^{2}\geq0\,,\nonumber \\
\mathfrak{q} & =2\omega g^{0}\left(p_{z}-2\zeta g^{0}n_{\chi}\right)\,,\nonumber \\
\mathfrak{r} & =\left[\left(2n_{\chi}\omega\right)^{2}+\left(p_{z}+\frac{\omega}{2}\right)^{2}-\left(\frac{g^{0}}{2}\right)^{2}\right]\left[\left(2n_{\chi}\omega\right)^{2}+\left(p_{z}+\frac{\omega}{2}\right)^{2}-\left(\frac{g^{0}}{2}\right)^{2}-\mathfrak{q}\right]\nonumber \\
 & +2m^{2}\left[\left(2\omega n_{\chi}+\zeta\frac{g^{0}}{2}\right)^{2}-\frac{\omega^{2}}{4}+p_{z}^{2}\right]+m^{4}\,.\label{8.2.4}
\end{align}
The solution of Eq.~(\ref{8.2.3}) can be obtained as follows. Using the new variable $\xi$, we rewrite Eq.~(\ref{8.2.3}) as
\begin{equation}
w^{4}+\mathfrak{p}w^{2}+\mathfrak{q}w+\mathfrak{r}=\left(w^{2}+\frac{\mathfrak{p}}{2}+\xi\right)^{2}-\left(2\xi w^{2}-\mathfrak{q}w+\frac{\mathfrak{p}^{2}}{4}-\mathfrak{r}+\mathfrak{p}\xi+\xi^{2}\right)=0\,.\label{8.2.5}
\end{equation}
In order for the trinomial $2\xi w^{2}-\mathfrak{q}w+\mathfrak{p}^{2}/4-\mathfrak{r}+\mathfrak{p}\xi+\xi^{2}$ to be reduced to a perfect square $(w^{2}+\mathfrak{p}/2+\xi)^{2}$, its discriminant should be equal to zero:
\begin{equation}
\xi^{3}+\mathfrak{p}\xi^{2}+\left(\frac{\mathfrak{p}^{2}}{4}-\mathfrak{r}\right)\xi-\frac{\mathfrak{q}^{2}}{8}=0\,.\label{8.2.6}
\end{equation}
If $\xi=\xi_{1}$ is some root of the cubic Eq.~(\ref{8.2.6}), then
\begin{equation}
2\xi w^{2}-\mathfrak{q}w+\frac{\mathfrak{p}^{2}}{4}-\mathfrak{r}+\mathfrak{p}\xi+\xi^{2}=2\xi_{1}\left(w-\frac{\mathfrak{q}}{4\xi_{1}}\right)^{2}\,.\label{8.2.7}
\end{equation}
Substituting Eq.~(\ref{8.2.7}) into Eq.~(\ref{8.2.5}), we obtain
\begin{equation}
\left(w^{2}+\frac{\mathfrak{p}}{2}+\xi_{1}\right)^{2}=2\xi_{1}\left(w-\frac{\mathfrak{q}}{4\xi_{1}}\right)^{2}\,,\label{8.2.8}
\end{equation}
then
\begin{equation}
w^{2}+\frac{\mathfrak{p}}{2}+\xi_{1}=\varepsilon_{1}\sqrt{2\xi_{1}}\left(w-\frac{\mathfrak{q}}{4\xi_{1}}\right),\quad\varepsilon_{1}=\pm1\,.\label{8.2.9}
\end{equation}
We are interested in the real roots of Eq.~(\ref{8.2.3}). Therefore, let $\xi_{1}$ be the positive root of cubic Eq.~(\ref{8.2.6}),
\begin{equation}
\xi_{1}=\sqrt[3]{-\frac{\mathfrak{q}}{2}-\sqrt{Q}}+\sqrt[3]{-\frac{\mathfrak{q}}{2}+\sqrt{Q}},\quad Q=\left(\frac{\mathfrak{p}}{3}\right)^{3}+\left(\frac{\mathfrak{q}}{2}\right)^{2}\,.\label{8.2.10}
\end{equation}
Using Eq.~\eqref{8.2.10}, one derives Eq.~\eqref{5.9} for the energy levels.


\begin{thebibliography}{50}

\bibitem{Fuk98}
  Y.~Fukuda et al. Super-Kamiokande Collaboration),
  Evidence for Oscillation of Atmospheric Neutrinos, 
  Phys. Rev. Lett. \textbf{81}, 1562--1567 (1998)
  [hep-ex/9807003].

\bibitem{Ahm02}
  Q.~R.~Ahmad et al. (SNO Collaboration),
  Direct Evidence for Neutrino Flavor Transformation from Neutral-Current Interactions in the Sudbury Neutrino Observatory,
  Phys. Rev. Lett. \textbf{89}, 011301 (2002)
  [nucl-ex/0204008].

\bibitem{Wol78}
  L.~Wolfenstein,
  Neutrino Oscillations in Matter,
  Phys. Rev. D \textbf{17}, 2369--2374 (1978).

\bibitem{MikSmi85}
  S.~P.~Mikheyev and A.~Yu.~Smirnov,
  Resonance Amplification of Oscillations in Matter and Spectroscopy of Solar Neutrinos,
  Sov. J. Nucl. Phys. \textbf{42}, 913--917 (1985).

\bibitem{CheXu25}
  S.~Chen and X.-J.~Xu,
  Solar neutrinos,
  arXiv:2501.09971.

\bibitem{GiuKim07}
  C.~Giunti and C.~W.~Kim, 
  \textit{Fundamentals of Neutrino Physics and Astrophysics}
  (Oxford University Press, Oxford, 2007).

\bibitem{Lor97}
  D.~R.~Lorimer, M.~Bailes, and P.~A.~Harrison,
  Pulsar statistics -- IV. Pulsar velocities,
  Mon. Not. R. Astron. Soc. \textbf{289}, 592--604 (1997).

\bibitem{Joh05}
  S.~Johnston, G.~Hobbs, S.~Vigeland, M.~Kramer, J.~M.~Weisberg, and A.~G.~Lyne,
  Evidence for alignment of the rotation and velocity vectors in pulsars,
  Mon. Not. R. Astron. Soc. \textbf{364}, 1397--1412 (2005)
  [astro-ph/0510260]
  
\bibitem{KusSer96}
  A.~Kusenko and G.~Segr\`{e},
  Pulsar Velocities and Neutrino Oscillations,
  Phys. Rev. Lett. \textbf{77}, 4872 (1996)
  [hep-ph/9606428].

\bibitem{LaiQia98}
  D.~Lai and Y.-Z.~Qian,
  Parity violation in neutrino transport and the origin of pulsar kicks,
  Astrophys. J. \textbf{495}, L103--L106 (1998)
  [astro-ph/9712043].

\bibitem{Vil80}
  A. Vilenkin,
  Equilibrium parity-violating current in a magnetic field,
  Phys. Rev. D \textbf{22}, 3080 (1980).

\bibitem{NieNin83}
  H.~B.~Nielsen and M.~Ninomiya,
  Adler-Bell-Jackiw anomaly and Weyl fermions in crystal,
  Phys. Lett. B \textbf{130}, 389 (1983).

\bibitem{SonSur09}
   D.~T.~Son and P.~Sur\'{o}wka,
  Hydrodynamics with Triangle Anomalies,
  Phys. Rev. Lett. \textbf{103}, 191601 (2009)
  [arXiv:0906.5044].

\bibitem{Kam16}
  M.~Kaminski, C.~F.~Uhlemann, M.~Bleicher, and J.~Schaner-Bielich,
  Anomalous hydrodynamics kicks neutron stars,
  Phys. Lett. B \textbf{760}, 170--174 (2016)
  [arXiv:1410.3833].

\bibitem{YamYan23}
  N.~Yamamoto and D.-L.~Yang,
  Effective Chiral Magnetic Effect from Neutrino Radiation,
  Phys. Rev. Lett. \textbf{131}, 012701 (2023)
  [arXiv:2211.14465].

\bibitem{Dvo16}
  M.~Dvornikov,
  Role of particle masses in the magnetic field generation driven by the parity violating interaction,
  Phys. Lett. B \textbf{760}, 406--410 (2016)
  [arXiv:1608.04940].

\bibitem{Dvo18a}
  M.~Dvornikov,
  Equilibrium electric current of massive electrons with anomalous magnetic moments induced by
  a magnetic field and the electroweak interaction with matter,
  Int. J. Mod. Phys. A \textbf{33}, 1850154 (2018)
  [arXiv:1801.07788].

\bibitem{Dvo18b}
  M.~Dvornikov,
  Chiral magnetic effect in the presence of an external axial-vector field,
  Phys. Rev. D \textbf{98}, 036016 (2018)
  [arXiv:1804.10241].

\bibitem{Dvo15}
  M.~Dvornikov,
  Galvano-rotational effect induced by electroweak interactions in pulsars,
  J. Cosmol. Astropart. Phys. \textbf{05}, 037 (2015)
  [arXiv:1503.00608].

\bibitem{Dvo14}  
  M.~Dvornikov, 
  Neutrino interaction with matter in a noninertial frame, 
  J. High Energy Phys. \textbf{10}, 053 (2014)
  [arXiv:1408.2735].

\bibitem{GriSavStu07}  
  A.~V.~Grigoriev, A.~M.~Savochkin, and A.~I.~Studenikin, 
  Quantum states of the neutrino in a nonuniformly moving medium, 
  Russ. Phys. J. \textbf{50}, 845--852 (2007).

\bibitem{BalPopStu11}  
  I.~Balantsev, Yu.~Popov, and A.~Studenikin, 
  On the problem of relativistic particles motion in strong magnetic field and dense matter, 
  J. Phys. A \textbf{44}, 255301 (2011)
  [arXiv:1012.5592].

\bibitem{Ver23}
  S.~N.~Vergeles, N.~N.~Nikolaev,  Yu.~N.~Obukhov, A.~Ya.~Silenko, and
  O.~V.~Teryaev,
  General relativity effects in precision spin experimental tests of
  fundamental symmetries,
  Phys. Usp. \textbf{66}, 109--147 (2023)
  [arxiv:2204.00427].
  
\bibitem{BagGit14book}
  V.~G.~Bagrov and D.~M.~Gitman, 
  \textit{The Dirac equation and its Solutions}
  (De Gruyter, Boston, 2014).

\bibitem{Dvo11}
  M.~Dvornikov,
  Field theory description of neutrino oscillations,
  in \textit{Neutrinos: Properties, Sources and Detection},
  ed. by J.~P.~Greene (Nova Science Publishers, New York, 2011),
  pp. 23-90
  [arXiv:1011.4300].

\bibitem{DvoStu02}
  M.~Dvornikov and A.~Studenikin,
  Neutrino spin evolution in presence of general external fields,
  J. High Energy Phys. \textbf{09}, 016 (2002)
  [hep-ph/0202113].
  
\bibitem{Collas2019}  
P.~Collas and D.~Klein,
\textit{The Dirac Equation in Curved Spacetime: A Guide for Calculations}
(Springer, Berlin, 2019).

\bibitem{Birrell1984}
N.~D.~Birrell and P.~C.~W.~Davies, 
\textit{Quantum fields in curved space}
(Cambridge University Press, Cambridge, 1984).  

\bibitem{BagGit90book}
V.~G.~Bagrov and D.~M.~Gitman, 
\textit{Exact solutions of relativistic wave equations}
(Dordrecht, Kluwer, 1990).

\bibitem{BagOb92}
V.~G.~Bagrov and V.~V.~Obukhov, 
New method of integration for the Dirac equation on a curved space-time,
J. Math. Phys. \textbf{33}, 2279 (1992).

\bibitem{Kha16}
  D.~E.~Kharzeev, J.~Liao, S.~A.~Voloshin, and G.~Wang,
  Chiral Magnetic and Vortical Effects in High-Energy Nuclear Collisions --- A Status Report,
  Progr. Part. Nucl. Phys. \textbf{88}, 1--28 (2016)
  [arXiv:1511.04050].

\bibitem{FlaFuk17}
  A.~Flachi and K.~Fukushima,
  Chiral vortical effect with finite rotation, temperature, and curvature,
  Phys. Rev. D \textbf{98}, 096011 (2018)
  [arXiv:1702.04753].
  
  \bibitem{Olivera25}  
  R.~R.~S.~Oliveira,
  Comment on ``Neutrino Interaction with Matter in a Noninertial Frame'',
  J. High Energy Phys. \textbf{01}, 085 (2025)
  [arXiv:2411.04338].
 
\bibitem{BrSh16} 
  A.~I.~Breev and A.~V.~Shapovalov,
  The Dirac equation in an external electromagnetic field: algebra and exact integration,
  J. of Phys.: Conf. Ser. \textbf{670}, 01201 (2016)
  [arXiv:1509.08612].

\bibitem{BagGitSmir04}  
  S~.P.~Gavrilov, D.~M.~Gitman and A.~A~.Smirnov, 
  Dirac equation in magnetic-solenoid field, 
  Eur Phys J C \textbf{32}, 119--142 (2004)


  
\end{thebibliography}
\end{document}